\documentclass{aastex61}

\usepackage[version=4]{mhchem}
\usepackage{chngcntr}
\usepackage{graphicx}
\usepackage{subfigure}
\received{}
\revised{}
\accepted{}
\submitjournal{ApJ}

\begin{document}

\title{The effect of carbon grain destruction on the chemical structure of protoplanetary disks }

\author{Chen-En Wei}
\affil{Department of Earth and Planetary Sciences, Tokyo Institute of Technology, 2-12-1 Ookayama, Meguro, Tokyo, 152-8551, Japan \\}

\author{Hideko Nomura}
\affil{Department of Earth and Planetary Sciences, Tokyo Institute of Technology, 2-12-1 Ookayama, Meguro, Tokyo, 152-8551, Japan \\}

\author{Jeong-Eun Lee}
\affiliation{School of Space Research, Kyung Hee University, Seocheon-Dong, Giheung-Gu, Yongin-Si, Gyeonggi-Do, 446-701, Republic of Korea}

\author{Wing-Huen Ip}
\affiliation{Graduate Institute of Astronomy, National Central University, No. 300, Zhongda Road, Zhongli Dist., Taoyuan City 32001, Taiwan}

\author{Catherine Walsh}
\affiliation{School of Physics and Astronomy, University of Leeds, Leeds, LS2 9JT, UK}

\author{T. J. Millar}
\affiliation{Astrophysics Research Centre, School of Mathematics and Physics, Queen's University Belfast, University Road, Belfast, BT7 1NN, UK}
\affiliation{Leiden Observatory, Leiden University, PO Box 9513, 2300 RA Leiden, The Netherlands}



\begin{abstract}

The bulk composition of the Earth is dramatically carbon poor compared to that of the interstellar medium, and this phenomenon extends to the asteroid belt. To interpret this carbon deficit problem, the carbonaceous component in grains must have been converted into the gas-phase in the inner regions of protoplanetary disks prior to planetary formation. We examine the effect of carbon grain destruction on the chemical structure of disks by calculating the molecular abundances and distributions using a comprehensive chemical reaction network. When carbon grains are destroyed and the elemental abundance of the gas becomes carbon-rich, the abundances of carbon-bearing molecules, such as HCN and carbon-chain molecules, increase dramatically near the midplane, while oxygen-bearing molecules, such as \ce{H2O} and \ce{CO2}, are depleted. We compare the results of these model calculations with the solid carbon fraction in the solar system. Although we find a carbon depletion gradient, there are some quantitative discrepancies: the model shows a higher value at the position of asteroid belt and a lower value at the location of the Earth. 
In addition, using the obtained molecular abundances distributions, coupled with line radiative transfer calculations, we make predictions for ALMA to potentially observe the effect of carbon grain destruction in nearby protoplanetary disks. The results indicate that HCN, \ce{H^{13}CN} and c-\ce{C3H2} may be good tracers. 


\end{abstract}




\section{Introduction} \label{sec:intro}

The protoplanetary disks (PPD) around low-mass stars are analogs of the solar nebula. Disks are composed of dust and gas surrounding a central pre-main-sequence star. As dust grains settle down to the midplane, they collide, grow and form planetesimals. These are the ingredients which form rocky planets, the cores of gas-giant planets and small bodies in our solar system. Therefore, the composition of rocky bodies contain information on the refractory materials available during the formation of solar system and provide clues on both the local and large-scale chemical and physical processes in disks.

Figure \ref{fig:1} shows that the abundance of carbon relative to silicon has clear variances across the solar system bodies (e.g., \citealt{1987A&amp;A...187..859G}; \citealt{2010ApJ...710L..21L}; \citealt{2014prpl.conf..363P}; \citealt{2015PNAS..112.8965B}; \citealt{2017ApJ...845...13A}; \citealt{2017MNRAS.469S.712B}). The similar amount of carbon relative to silicon in the Sun and ISM suggests that this value is representative of the ratio at the beginning of the formation of the planets. Although the surface of our Earth is covered with organic compounds, including even live organisms, if one excludes the uncertain amount of carbon in the core, the Earth BSE value (Bulk Silicate Earth: includes only ocean, atmosphere and silicate mantle) is 4 orders of magnitude lower than the solar ratio. Even when taking into account that some light elements, including carbon, are possibly incorporated inside the core of Earth, the abundance ratio remains significantly lower than solar (\citealt{2001E&amp;PSL.185...49A}). In the asteroid belt, the amount of carbon in meteorites relative to silicon is 1 to 2 orders of magnitude less than the solar value. Moving farther out, to the comet region, the carbon fractions become again comparable to that in the Sun. The C/Si ratio shows a clear correlation with heliocentric distance. Highly depleted elements in the solar system imply that they were in a phase too volatile to condense, and as such, they were not incorporated into planetesimals. Thus any gaseous carbon species which did not freeze onto the surface of grains nor become incorporated into refractory organic material will be either accreted onto the central star or dispersed out of the disk. In contrast, refractory carbon or carbon contained in ice mantles would have been incorporated into solid bodies producing, for example, the solar-like carbon abundance relative to silicon seen in comets.

The described trend of carbon depletion in the solar system is a long-standing problem in the community (\citealt{1987A&amp;A...187..859G}; \citealt{2010ApJ...710L..21L}; \citealt{2014prpl.conf..363P}; \citealt{2015PNAS..112.8965B}; \citealt{2017ApJ...845...13A}). This is the so-called carbon deficit problem. This issue exists not only in the solar system but also in white dwarf systems (\citealt{2006ApJ...653..613J}). Therefore, the cause might be universal throughout the Galaxy. In the ISM about 60\% of carbon is in the form of either graphite, amorphous carbon grains, or polycyclic aromatic hydrocarbons (PAHs) (\citealt{1996ARA&amp;A..34..279S}; \citealt{2003ARA&amp;A..41..241D}). If the solar system inherited ISM-like material, then refractory carbon must have been destroyed prior to planetary formation to account for the observed carbon depletion. Some possible ways to destroy carbon grains have been discussed: (1) carbon can be released from the solid to the gas phase by hot oxygen atoms, if the gas temperature is higher than 500 K and the grain size is within 0.1-1 $\micron$ (\citealt{2010ApJ...710L..21L}). (2) carbon aggregates such as graphite can react with oxygen-bearing species, e.g., OH (\citealt{1997A&amp;A...325.1264F}; \citealt{2001A&amp;A...378..192G}; \citealt{2002A&amp;A...390..253G}). (3) PAH destruction by X-ray and extreme ultraviolet (EUV) photons (\citealt{2010A&amp;A...511A...6S}). 

We start by considering the nature of a PPD, models of which are complex due to the interaction between physical and chemical processes of the dust and gas. The environment of the protoplanetary disk is diverse. The density varies by about 10 orders of magnitude and the temperature ranges from a few K to thousands of K. From the surface to the midplane, the main structure can be classified into three layers; a surface PDR (Photon Dominated Region)-like layer, a warm molecular layer and a cold midplane. The PDR-like layer mainly consists of atoms and ions due to the severe stellar UV radiation and interstellar radiation field the latter of which can influence the disk structure at larger radial distances from the star. In this layer, chemistry induced by photodissociation and photoionization dominates. Moving downward to the warm molecular layer, molecules here can survive due to partial shielding from the radiation field. In this layer, gas phase two-body reactions (neutral-neutral reactions and ion-neutral reactions) are active and set the molecular composition. Towards the midplane, excepting the very inner part, the temperature decreases due to the efficient shielding of radiation. Therefore, in the midplane, molecules can freeze on the surfaces of dust grains. Chemical modeling of PPDs has been widely described in, for example, \citet{2007prpl.conf..751B}, \citet{2013ChRv..113.9016H} and \citet{2014prpl.conf..317D}. 

The distribution of molecules is important for planetary formation, especially near the midplane. The final composition of planets is related to the location of snowlines in comparison with the planet's formation location. The balance between desorption and adsorption of gas-phase species onto grains determines the composition of the matter forming planetesimals and planets. From the observational point of view, several molecules, such as CO, \ce{CO2}, CN, HCN, \ce{H2CO}, \ce{C2H2}, OH (e.g., \citealt{1997A&amp;A...317L..55D}; \citealt{2004A&A...425..955T}; \citealt{2010ApJ...720..887P}) and some deuterated molecules, \ce{DCO+}, and DCN (e.g., \citealt{2003A&amp;A...400L...1V}; \citealt{2008ApJ...681.1396Q}) have been detected in the disks so far. Recently, some relatively complex molecules in PPDs have been detected, such as \ce{CH3CN}, \ce{CH3OH} and \ce{HCOOH} (\citealt{_berg_2015}; \citealt{Walsh_2016}; \citealt{Favre_2018}; \citealt{Loomis_2018}; \citealt{Bergner_2018}). So far, the total molecular inventory in PPDs is $>$ 20. Thanks to the high spatial resolution and sensitivity of ALMA, we have opportunities to reveal further detail, even in the inner region of disks ($<$ 10's of au).

In this paper, we calculate the chemistry in protoplanetary disks taking into account carbon grain destruction in the inner region to study how it affects the molecular abundance distribution in the disk. We compare the results between cases with and without carbon grain destruction, finding the candidate species which have the largest variations in abundance between two cases. To investigate whether or not the effect of carbon grain destruction is observable with ALMA, we perform the radiative transfer calculations of molecular line emission from identified species in T Tauri and Herbig Ae disks.

We describe our disk model and the radiative transfer calculations in  Section \ref{sec:2}. The results and discussion are presented in Section \ref{sec:3}: Section \ref{subsec:3_1} presents the effect of carbon grain destruction on the molecular abundances and distributions (HCN, \ce{CH4}, \ce{C2H2}, \ce{c-C3H2} carbon-chain molecules, \ce{H2O}, OH, \ce{O2} and \ce{CO2}). Section \ref{subsec:3_2} discusses the radial distribution of the solid carbon fraction in the disk in our model in comparison with the solar system, and Section \ref{subsec:3_3} presents the prediction for ALMA observation based on our calculations. We summarize and state the conclusions of this work in Section \ref{sec:4}.

\section{ Protoplanetary disk model}  \label{sec:2}

\subsection{Physical model} \label{subsec:2_1}

We use physical models of protoplanetary disks generated using the methodology described in \citet{2005A&amp;A...438..923N} with the addition of X-ray heating as described in \citet{2007ApJ...661..334N}. We model an axisymmetric Keplerian disk and two types of central stars are considered. The first is a typical T Tauri star with mass, $M_\ast$ = 0.5  $M_\sun$, radius, $R_\ast$= 2.0 $R_\sun$, and effective temperature, $T_\ast$= 4000 K (\citealt{1995ApJS..101..117K}). The second is a Herbig Ae star with mass, $M_\ast$= 2.5 $M_\sun$, radius, $R_\ast$= 2.0 $R_\sun$, and effective temperature, $T_\ast$= 10,000 K (e.g., \citealt{1993ApJ...418..414P}). 
	
The gas density profiles are determined by assuming vertical hydrostatic equilibrium, the balance between gravity and the pressure gradient force. In order to obtain the gas surface density profiles, we adopt the viscous $\alpha$-disk model (\citealt{1973A&amp;A....24..337S}), with a viscous parameter, $\alpha=10^{-2}$ and a mass accretion rate, $\dot{M} = 10^{-8}  M_\sun \mbox{yr}^{-1}$. The temperature profile of the gas is calculated self-consistently with the gas density profiles by assuming local thermal balance between gas heating and cooling. The thermal processes of heating from X-ray ionization of hydrogen, grain photoelectric heating induced by far-ultraviolet photons and cooling via gas-grain collisions and line transitions are taken into account. The dust temperature profiles are obtained by assuming local radiative equilibrium between the blackbody emission of grains and the absorption of radiation from the central star as well as the surrounding dust grains.
	
Regarding dust, we assume that the dust and gas are well-coupled and that the dust-to-gas mass ratio is constant (0.01) throughout the disk. The dust size distribution model in \citet{2001ApJ...548..296W}, which reproduces the observational extinction curve of dense clouds, is adopted. The dust properties are important factors for both gas and dust temperatures. The adopted dust opacity is described in Appendix D of \citet{2005A&amp;A...438..923N}. 
	
For the UV radiation field two radiation sources are taken into account: radiation from the central star and the interstellar medium. In the case of the T Tauri star, we use a UV excess model that reproduces the observational data toward TW Hydrae. The components of the stellar UV radiation are blackbody emission, hydrogenic thermal Bremsstrahlung emission and Ly$\alpha$ line emission (see Appendix C of \citealt{2005A&amp;A...438..923N} for details). For the Herbig Ae star, the UV excess contributes a significantly smaller fraction to the total UV luminosity compared with that from the stellar blackbody radiation (e.g., \citealt{2013PASP..125..477D}). Therefore, we use only the blackbody emission of the central star as the source of UV radiation field for the Herbig Ae model. For the X-ray radiation, we adopt a TW-Hydrae-like spectrum with a total luminosity, $L_X \sim 10^{30} \mbox{ erg s}^{-1}$, for the T Tauri star (\citealt{2005ApJS..160..401P}). Due to the non-convective interiors of Herbig Ae stars, the typical X-ray luminosities are $\ga$ 10 times lower than those of T Tauri stars (\citealt{2009A&amp;ARv..17..309G}). Therefore, we assume an X-ray spectrum with $L_X  \sim 3 \times 10^{29}  \mbox{ erg s}^{-1}$ and $T_X \sim 1.0 \mbox{ keV}$  for the Herbig Ae star based on observations (e.g., \citealt{1994A&amp;A...292..152Z}; \citealt{2005ApJ...618..360H}). 

For full details on the generation of the physical models, see \citet{2005A&amp;A...438..923N} and \citet{2007ApJ...661..334N}.
The total grid numbers are 8699 and 12116 for the disks around T Tauri and Herbig Ae stars, respectively, where 129 radial steps are taken logarithmically for the disk radius from 0.04 to 305 au. Figure \ref{fig:2} shows the physical structure of the disk around the T Tauri star and the Herbig Ae star (see also \citealt{2015A&amp;A...582A..88W}; {\citealt{2016ApJ...827..113N, 2017ApJ...836..118N}). From top, the gas number density decreases as the function of height and the radius. The densest region reaches about $10^{15} \mbox{cm}^{-3}$ near the disk midplane, falling to $10^5 \mbox{cm}^{-3}$ in the surface layer, showing the diversity of the environment in the PPDs. The gas and dust temperature decouple in the diffuse region due to inefficient collisions between the gas and dust grains. As the density increases toward the disk midplane, the gas and dust temperature become well coupled. The UV flux from the central star is much stronger in the case of Herbig Ae star (right-hand-side) than in the T-Tauri star (left-hand-side). However, both UV and X-ray photons are absorbed by dust and gas and do not penetrate to the midplane. Since the Herbig Ae star has a higher luminosity, its enhanced gas and dust temperatures make line emission easier to detect than in the T Tauri case (see Section \ref{subsec:3_3}). Noted that different radial ranges are shown in Figure \ref{fig:2} for each type of disk.

\subsection{Chemical reaction network} \label{subsec:2_2}

We use a chemical reaction network, containing both gas-phase reactions and gas-grain interactions to study the chemical evolution of the disks. In order to examine the effect of carbon grain destruction (\citealt{2010ApJ...710L..21L}; \citealt{2017ApJ...845...13A}), we focus on those small molecules that are not much affected by grain surface reactions and consider two initial values for the elemental abundance of carbon. Except the abundance of carbon, for both models without and with carbon grain destruction, we use the typical low-metal value often adopted to model the chemistry of dark clouds such as TMC-1 (Table 8 of \citealt{2007A&amp;A...466.1197W}; see also Table \ref{tab:1}). In the model without carbon grain destruction, the elemental abundance of carbon in gas is the same as that in diffuse clouds, in which the gas-phase elemental abundance of carbon is lower than that of oxygen. For the model with carbon grain destruction, we consider an extreme case assuming that all the carbon in the grains is destroyed and released into the gas-phase, and set the abundance for ionized carbon to be the solar abundance of carbon, $\ce{C+}=2.95 \times 10^{-4}$ (e.g., \citealt{2009ARA&amp;A..47..481A}; \citealt{draine2010physics}). In this case, the gas-phase elemental abundance of carbon is larger than that of oxygen. We note that the C/O ratio is larger than unity in the gas-phase in this extreme case, although the C/O ratio of the solar photosphere is around 0.5 \citep{2002ApJ...573L.137A}. This is because a large fraction of available elemental oxygen is incorporated into silicate grains which are more difficult to destroy than carbon grains. According to \cite{2010ApJ...710L..21L}, if the grain size is small enough, carbon grains could be sputtered into the gas phase in the surface layer of the disks where the gas temperature is high. If turbulent mixing in the vertical direction of the disk is efficient enough, a significant amount of the small carbon grains could be sputtered inside a disk radius of around 10 au in the case of T Tauri disks. Therefore, in this paper we only consider chemistry in the very inner region and treat the T Tauri disk up to 10 au, and the Herbig Ae disk, which the UV irradiation is stronger, up to 50 au. We calculate the chemical evolution up to $10^{6}$ years, which is the typical life time for protoplanetary disks and at which point the chemical structure has almost reached steady state in the disk. We note that for a more realistic model, the evolution of physical conditions, such as density, temperature, and radiation field, should be treated together with the time evolution of molecular abundances. This is beyond the scope of this work, and for simplicity, we use the chemical composition of diffuse clouds as an initial condition, and use the fixed physical conditions of each protoplanetary disk for the calculation of chemical reactions. Also, and again for simplicity, we ignore the change in the dust mass and surface area caused by the carbon grain destruction and the freeze-out of gas-phase molecules on grains during the calculation of chemical reactions.

\begin{table}[ht]
\centering
\begin {tabular} {c c c c c}
\hline
\hline
Element &   & Abundance &   &\\[0.5ex] 
\hline
 &  Without C grain destruction &  &  With C grain destruction &   \\[0.5ex] 
\hline
\ce{He}   &  &0.14  & & \\                
\ce{C^+}  & $7.30 \times 10^{-5}$ & & $2.95 \times 10^{-4}$ \\
\ce{N}    &  & $2.14 \times 10^{-5}$ & \\
\ce{O}    & & $1.76 \times 10^{-4}$ &\\
\ce{Na^+} & & $3.00 \times 10^{-9}$ & \\
 \ce{Mg^+} & & $3.00 \times 10^{-9}$ \\
  \ce{Si^+} & & $3.00 \times 10^{-9}$ \\
  \ce{S^+}  & & $2.00 \times 10^{-8}$ \\
   \ce{Cl^+} & & $3.00 \times 10^{-9}$ \\
   \ce{Fe^+} & & $3.00 \times 10^{-9}$ \\
\hline
\end{tabular}
\caption{The initial gas-phase elemental abundances relative to total hydrogen nuclei in the model without carbon grain destruction. For the model with carbon grain destruction, we adopt a carbon abundance of \ce{C+}=$2.95 \times 10^{-4}$.  }
\label{tab:1}
\end{table}


\subsubsection{ Gas-phase reactions} \label{subsec:2_3}

We use the RATE06 release of the UMIST Database for Astrochemistry, for the gas-phase chemistry (\citealt{2007A&amp;A...466.1197W}). In this chemical network, there are 375 species and 4336 reactions, including 3957 two-body reactions, 214 photoreactions, 154 X-ray/cosmic-ray-induced photoreactions, and 11 reactions of direct X-ray/cosmic-ray ionization. Since the reactions including fluorine, F, and phosphorus, P, have a low impact on the chemistry (\citealt{2010ApJ...722.1607W}), we remove reactions containing these two elements to reduce the computation time. For the photoreactions, we calculate the UV radiation at each point in the disk as 

\begin{equation} \label{eq:eq1}
G_{FUV}(r,z)=\int_{912 \AA (13.6 eV)}^{2068 \AA (6 eV)} G_{FUV}(\lambda,r,z)d\lambda.
\end{equation}
The UV radiation is scaled by the interstellar UV flux, $G_0=2.67 \times 10^{-3}  \mbox{ erg cm}^{-2} \mbox{s}^{-1}$ (\citealt{2006FaDi..133..231V}), and photoreaction rates at each point in the disk can be approximated as 

\begin{equation} \label{eq:eq2}
k^{ph}(r,z)=\frac{G_{FUV}(r,z)}{G_0}k_0 \quad s^{-1}, 
\end{equation}
where $k_0$ is the unshielded photoreaction rates due to the interstellar UV radiation field as compiled in RATE06 (see also \citealt{1997A&amp;AS..121..139M}).

\subsubsection{ Gas-grain interaction} 

The gas-grain interactions included are the adsorption (freeze-out) and desorption of molecules on and from dust grain surfaces, respectively. When the gas temperature is low enough, the adsorption rate becomes larger than the desorption rate, and therefore, molecules freeze onto dust grains. Otherwise, molecules sublimate from dust grains and into the gas-phase, so-called ``thermal desorption''. The rate of thermal desorption, and thus the temperature at which sublimation happens, will depend on the binding energy of each molecule to dust grains. The binding energies of several important molecules and adopted in their work are listed in Table \ref{tab:2}. In addition to thermal desorption, we consider non-thermal desorption mechanisms: cosmic-ray-induced thermal desorption (\citealt{1985A&amp;A...144..147L}; \citealt{1993MNRAS.263..589H}) and UV photodesorption (\citealt{1995P&amp;SS...43.1311W}; \citealt{2000ApJ...544..903W}; \citealt{2007ApJ...662L..23O}). The details of reaction rates for these processes can be found in the Appendix \ref{appen_a} (see also \citealt{2010ApJ...722.1607W}; \citealt{2016ApJ...827..113N}).

Considering the processes of freeze out, thermal desorption, cosmic-ray-induced desorption and photodesorption, the differential equation for the number density of species $i$ on the dust grain is written as (e.g., \citealt{1992ApJS...82..167H})

\begin{equation}\label{eq:eq8}
\frac{dn_{i,ice}}{dt}=n_ik_i^a-n_{i,ice}^{desorb}(k_i^d+k_i^{crd}+k_i^{pd}),
\end{equation}
where $k_i^a$ is the adsorption rate, $k_i^d$ is the thermal desorption rate, $k_i^{crd}$ represents the cosmic-ray-induced thermal desorption rate, and $k_i^{pd}$ denotes the photodesorption rate of species $i$. $n_i$ is the gas phase number density of species $i$ and $n_{i,ice}^{desorb}$ is the number density of species $i$ located in uppermost active layers of the ice mantle. The value of $n_{i,ice}^{desorb}$ is given by \citep{1996ApJ...467..684A} as 
\begin{equation}\label{eq:eq9}
{ n }_{ i,ice }^{ desorb }=\begin{cases} n_{i,ice} \quad \quad \quad(n_{ice}<n_{act})\\n_{act}\frac{n_{i,ice}}{n_{ice}} \quad (n_{ice}\ge n_{act})\end{cases}
\end{equation}
where $n_{ice}$ is the total ice number density of all species (see also \citealt{Walsh_2014}) and $n_{act}$=$4 \pi r_d^2n_dn_sN_{LAY}$ represents the number of active surface sites in the ice mantle per unit volume and $N_{LAY}$ is the number of surface layers considered as active, assumed to be two. $r_d$ is the dust grain radius, $n_s$ represents the surface density of sites and $n_d$ is the number density of dust grains. For more detail, please refer to Appendix \ref{appen_a}.

\subsection{ Radiative transfer and the line emission}

In order to predict observational signatures of carbon grain destruction in the inner region of protoplanetary disks, we perform ray-tracing calculations to estimate the intensity of molecular line emission and obtain intensity maps for both models with and without carbon grain destruction. The line data are taken from the Leiden Atomic and Molecular Database (LAMDA, \citealt{2005A&amp;A...432..369S}), the Cologne Database for Molecular Spectroscopy (CDMS, \citealt{M_ller_2005}), or the Jet Propulsion Laboratory (JPL) molecular spectroscopic database (\citealt{PICKETT_1998}).
We modify the original 1D code, RATRAN (\citealt{2000A&amp;A...362..697H}) to calculate the ray-tracing using an axisymmetric 2D disk structure under the assumption of local thermodynamic equilibrium (LTE).

The intensity of the line profile at a frequency $\nu$, $I_{ul}(\nu)$, is obtained by solving the radiative transfer equation along the line-of-sight $s$ in the disk
\begin{equation}\label{eq:eq10}
\frac{dI_{ul}(\nu)}{ds}=-\chi_{ul}(\nu)(I_{ul}(\nu)-S_{ul}(\nu))
\end{equation}
where $\chi_{ul}(\nu)$ is the total extinction coefficient of dust grains and molecular lines and $S_{ul}(\nu)$ is the source function given by 

\begin{equation}\label{eq:eq11}
\chi_{ul}(\nu)=\rho_d\kappa_{ul}+\frac{h\nu_{ul}}{4\pi}(n_lB_{lu}-n_uB_{ul})\Phi_{ul}(\nu)
\end{equation}
and
\begin{equation}\label{eq:eq12}
S_{ul}(\nu)=\frac{1}{\chi_{ul}(\nu)}\frac{h\nu_{ul}}{4\pi}n_{u}A_{ul}\Phi_{ul}(\nu)
\end{equation}
respectively, where $A_{ul}$ and $B_{ul}$ are the Einstein A and B coefficients for spontaneous and stimulated emission and $B_{lu}$ is the Einstein B coefficient for absorption. $n_u$ and $n_l$ are the number densities of the molecules in the upper and lower levels, respectively.
$\rho_d$ is the mass density of dust grains where we simply apply the dust to gas mass ratio of $\rho_d/\rho_g=1/100$. $\kappa_{ul}$ is the dust absorption coefficient at a frequency $\nu_{ul}$ and $h$ is Plank's constant. 

Here $\Phi(\nu)$ is the line profile function which is affected by thermal broadening and Keplerian rotation in the disk. As a result, $\Phi(\nu)$, is given by

\begin{equation}\label{eq:eq13}
\Phi_{ul} (\nu )=\frac { 1 }{\Delta \nu_D \sqrt{\pi}} \exp \left[ \frac{-(\nu+\nu_K-\nu_{ul})^2}{\Delta\nu_D^2} \right],
\end{equation}
where $\Delta \nu _D= (\nu_{ul}/c)\sqrt{2k_BT_g/m_i }$ is the Doppler width, $c$ is the speed of light,  $T_g$ represents the temperature of gas and $m_i$ is the mass of the species, $i$. $\nu_K$ denotes the Doppler shift due to the projected Keplerian velocity along the line-of-sight and is given by   
\begin{equation}\label{eq:eq14}
\nu _K=\frac{\nu_{ul}}{c}\sqrt{\frac{GM_*}{r}}\sin\phi \sin\theta,
\end{equation}
where $G$ is the gravitational constant, $M_{\ast}$ is the mass of the central star, $r$ is the distance of the line emitting region from the central star, $\phi$ is the azimuthal angle between the semimajor axis and the line linking the point in the disk along the line of sight and the center of the disk, and $\theta$ is the inclination angle of the disk. 
	
In order to predict the observable flux density, we integrate Equation \ref{eq:eq10} along the line of sight $s$. The intensity at $(x,y)$ in the projected plane is given by
\begin{equation}\label{eq:eq15}
I(x,y,\nu)=\int_{-s_\infty}^{s_\infty}{j_{ul}(s,x,y,\nu)}\exp(-\tau _{ul}(s,x,y,\nu))ds,
\end{equation}
where $j_{ul}$ is the emissivity at the position $(s,x,y)$ and frequency $\nu$, given by
\begin{equation}\label{eq:eq16}
j_{ul}(s,x,y,\nu)=n_u(s,x,y)A_{ul}\frac{h\nu_{ul}}{4\pi}\Phi_{ul}(s,x,y,\nu),
\end{equation}
and $\tau_{ul} (s,x,y,\nu)$ is the optical depth from the line emitting point $s$ to the disk surface $s_\infty$ at the frequency $\nu$, expressed by
\begin{equation}\label{eq:eq17}
\tau_{ul}(s,x,y,\nu)=\int_{s}^{s_\infty} \chi_{ul}(s',x,y,\nu)ds'.
\end{equation}
The observable line flux integrated all over the disk is given by
\begin{equation}\label{eq:eq18}
  F_{ul}(\nu)=\frac{1}{4 \pi d^2}\iint{I(x,y,\nu)dxdy}, 
\end{equation}
where $d$ is the distance between the observer and the target object.

\section{ Result and Discussion}  \label{sec:3}

\subsection{Effect of carbon grain destruction on molecular abundance distribution} \label{subsec:3_1}

In this subsection, we show effect of carbon grain destruction on carbon-bearing species (HCN, \ce{CH4}, \ce{C2H2}, carbon-chain and hydrocarbon molecules) and oxygen-bearing species (\ce{H2O}, OH, \ce{O2} and \ce{CO2}), some of which have been detected in protoplanetary disks at infrared wavelengths (e.g., \citealt{2004ApJ...603..213C}; \citealt{2006ApJ...636L.145L}; \citealt{2008Sci...319.1504C}; \citealt{2008ApJ...676L..49S}; \citealt{2009ApJ...696..143P}; \citealt{2007ApJ...660.1572G}). Here we mainly focus on the T Tauri disk model as an analogue of our solar system.

First, in order to see how the physical properties affect the molecular abundance profiles, the top panels of Figure \ref{fig:3} show the number density of hydrogen nuclei, the gas temperature and dust temperature as a function of height divided by radius (Z/r) at a disk radius of 1 au (left), 3 au (middle) and 10 au (right). The bottom panels show similar plots for the UV flux, the X-ray ionization rate and the cosmic-ray ionization rate.
	
Since gas-phase molecular abundances change dramatically across their snowlines, we estimate the location of the snowlines of some molecules by equating the adsorption rate (Equation \ref{eq:a1}) with the desorption rate (Equation \ref{eq:a2}). Because thermal desorption rate is related to the binding energy, Table \ref{tab:2} presents the location of the snowline of each species together with binding energies adopted in the model.

\begin{table}[h]
\centering
\begin{tabular}{l c  c c}
\hline
\hline
Molecule & Binding Energy (K) & Snowline in T-Tauri Disk (au)  & Snowline in HAe Disk (au) \\
\hline
CO                 & 855    & $>$10  & $>$50 \\
CH$_4$         & 1080    & $>$10  & $>$50 \\
C$_2$H$_2$ & 2400    & 9.33     & $>$50 \\
CO$_2$        & 2990     & 5.34     & 37.65  \\
C$_5$           & 3220     & 2.66    & 32.75  \\
C$_6$           & 3620    & 2.48     & 24.77  \\
C$_6$H        & 3880    & 2.31     & 21.54  \\
HCN             & 4170    & 2.31     & 18.74  \\
C$_7$          & 4430    & 2.15     & 17.48  \\
C$_8$          & 4830    & 2.01     & 14.18  \\
H$_2$O       & 4820    & 2.01     & 14.18  \\
HCOOH      & 5000     & 2.01     & 13.22  \\
C$_9$         & 5640    & 1.75     & 10.00  \\
C$_{10}$     & 6000    &1.52      & 8.70   \\

\hline
\end{tabular} 
\caption{The locations of the snowlines in T-Tauri and Herbig Ae disks determined by the temperature profiles given in Figure \ref{fig:2}.}
\label{tab:2}
\end{table}

Figures \ref{fig:4}, \ref{fig:6}, \ref{fig:8} and \ref{fig:10} show molecular abundances relative to total hydrogen nuclei in both the gas-phase and ice mantle at $10^6$ years as a function of (Z/r) at the disk radii of 10 au, 3 au and 1 au.
The dashed and the solid lines represent the case without and with carbon grain destruction, respectively. Figures \ref{fig:5}, \ref{fig:7}, \ref{fig:9} and \ref{fig:11} display the 2-dimensional fractional abundances of gas-phase and ice mantle molecules as a function of radius and height, out to a maximum radius of 10 au since the destruction of carbon grains mainly occurs in the inner hot region only. In the latter set of figures, the left-hand-side plots represent the case without carbon grain destruction (C$/$O$<$1) and right-hand-side plots those with carbon grain destruction (C$/$O$>$1).  

Since the abundance of gas-phase CO has a large effect on molecular abundances differences between the models with and without the carbon grain destruction, as is shown below, we present first the gas-phase and ice mantle abundances of CO (Figure \ref{fig:4}). In the disk surface, CO gas is dissociated by the FUV photons so that oxygen and carbon are not locked in CO but are mostly in the form of atoms and ions since the synthesis of molecules is difficult due to the high flux of FUV photons. Figure \ref{fig:5} shows the global CO abundance distribution, and CO is abundant (x(CO)=$10^{-4}$) throughout the whole region near the midplane, since it is easily desorbed inside 10 au due to its low binding energy. The abundance of CO increases slightly when carbon grains are destroyed, due to the additional elemental carbon now available in the gas phase. 

The abundances of other molecules, however, are significantly affected by the gas-phase elemental C/O abundance ratio, especially in the region where CO gas is abundant. In general, carbon-bearing molecules are not expected to be very abundant in an oxygen-rich environment (C/O$<$1) since most of the carbon is incorporated into CO. However, in the carbon grain destruction regions, excess carbon exists (C/O$>$1) and carbon-bearing molecular abundances can increase dramatically. This type of chemistry is similar to that observed and successfully modeled around carbon-rich AGB stars, such as IRC+10216 (e.g., \citealt{1994A&amp;A...288..561M}; \citealt{1996ARA&amp;A..34..241G}).

Figure \ref{fig:6} shows 1D plots of the abundances of some carbon-bearing species (HCN, \ce{CH4}, \ce{C2H2} and c-\ce{C3H2}). The left and right panels show the molecular abundances in gas and ice, respectively. Figure \ref{fig:7} shows the 2-D abundance distribution of the carbon-bearing species. The effects of carbon grain destruction, and the resulting chemistry, are different between the surface and the midplane. Near the midplane, carbon-bearing molecules can form efficiently via gas-phase reactions in the carbon-rich case. The most significant difference between two cases appears in HCN especially inside $\sim$ 2 au, near the HCN snowline.
Due to its high binding energy (see Table \ref{tab:2}), HCN can remain on dust grains beyond 2 au in the T Tauri disk. 
In the carbon-rich case, the peak gas-phase abundance of HCN is $\sim$$10^{-5}$ near the midplane of the inner region. Near the midplane inside 2 au, the maximum differences between two cases can reach 8 orders of magnitude. In the oxygen-rich case, the peak abundance is $\sim$ $10^{-7}$ in the molecular layer. Therefore, HCN appears to be a good tracer of the effect of carbon grain destruction within its snowline, and we will focus on it in section \ref{subsec:3_3}. We note, however, that the abundance distribution of HCN could be affected by the initial abundance of species and the ingredients adopted in the chemical model (see e.g., Figure 14 of \citealt{2015A&amp;A...582A..88W}).

Similar behavior can be seen in \ce{CH4} and \ce{C2H2}, but the differences are less significant. The snowlines of \ce{CH4} and \ce{C2H2} are located outside 10 au and  around 9.3 au, respectively. The gas phase abundances of both molecules increase near the midplane inside the snowlines in the carbon-rich case. However, inside 2 au, most of the carbon is incorporated into HCN and long carbon-chain species in the gas-phase (see Figure \ref{fig:8} and Figure \ref{fig:9}) and the abundances of \ce{CH4} and \ce{C2H2} gas decrease here. We note that \ce{CH4} is more abundant than HCN and carbon-chain molecules in the gas-phase near the midplane inside 2 au in the oxygen-rich case. In the oxygen-rich case \ce{CH4} gas has its peak abundance ($\sim$ $10^{-7}$) in the molecular layer.
In the case of c-\ce{C3H2}, it shows a differences of about 7 orders of magnitude between two cases. Though it is less abundant than HCN, we treat c-\ce{C3H2} as a possible tracer in section \ref{subsec:3_3} as well. We note that though the reaction network used here treats c-\ce{C3H2} and l-\ce{C3H2} separately, we should be cautious that there are some uncertainties since it is rarely possible to identify the specific isomeric products in a laboratory reaction and we need to rely on some calculations about energetics of the product and so on. Nevertheless, the model calculations show quite good agreement with observations of c- and l-\ce{C3H2} in IRC+10216 and TMC-1 (\citealt{2013A&A...550A..36M}).

Figure \ref{fig:8} presents the 1-D plots of \mbox{C$_m$($m=5-10$)}, \mbox{C$_m$H}($m=5-9$) and \mbox{C$_m$H$_2$}($m=5-9$) molecules, where we sum up the abundances of the molecules for $m=5-10$ or $m=5-9$. Figure \ref{fig:9} shows 2-D gas and ice distributions of \ce{C6H} and \ce{C9}. We chose these two species as representative of carbon-chain molecules since they are abundant on grain surfaces (Table \ref{tab:4}) and, indeed, most of the carbon-chain molecules have spatial distributions similar to these two species and to HCN. We note that we treat the carbon-chain molecules together since grain surface reactions, which may reduce the abundances of carbon-chain molecules, for example through hydrogenation, are not included here. From our calculations, we find that carbon-chain molecules with larger number of carbon atoms have larger abundances because the destruction of carbon-chain molecules by molecular ions is inefficient near the midplane of the disk where the density is high and the ionization degree is very low ($\sim$$10^{-12}$). In the carbon-rich case, the gas-phase abundances increase inside the snowline near the midplane. Beyond the snowline, the ice-mantle abundances increase significantly. The binding energies of carbon chain molecules are relatively large and thus the molecules can remain on the dust grains until very close to the parent star ($\sim$ 1.5 - 3 au).

We note that the gaseous carbon-chain molecules are abundant in the molecular layers beyond their snowlines in both cases. This region corresponds to the CO gas depletion layer. Figure \ref{fig:4} and Figure \ref{fig:5} show CO gas is depleted at $Z/r \sim 0.15, r = 3 - 10$ au. Similar CO gas depletion can be seen in other disk chemical models (e.g., \citealt{2010ApJ...722.1607W}; \citealt{2014ApJ...790...97F}). Though the self and mutual shielding of CO photodissociation are not included in the model, the shielding factors are not so effective especially in the inner region of the disks (\citealt{Walsh_2012}). In our model, the decrease is caused by photodissociation of CO due to FUV irradiation from the central star, and the subsequent formation and freeze-out of carbon-chain molecules onto grains. If the FUV field is strong enough, FUV photons dissociate CO into atomic species. Above the CO gas depletion area, the UV destruction is fast, and due to the low density, the adsorption rate on to grains is low. Therefore, any synthesized molecules are difficult to stick on the grains and will eventually convert to CO gas or, nearer the surface remain as atomic and ionized carbon. Moving deeper towards the disk midplane, where UV photons are extinguished by grains in the high density molecular layers, CO can remain in the gas-phase.
A fraction of the carbon-chain molecules in ice mantle in the CO gas depleted layer is photodesorbed into gas, which produces the gaseous carbon-chain molecule rich layer seen in Figure \ref{fig:8} and Figure \ref{fig:9}.

Figure \ref{fig:10} shows the 1-D plots of oxygen-bearing species (\ce{H2O}, \ce{OH}, \ce{O2} and \ce{CO2}) in the gas phase and ice mantle. Figure \ref{fig:11} shows the 2-dimensional distribution of gas-phase oxygen-bearing species, \ce{H2O} and \ce{CO2}. The results are opposite in behavior to those of carbon-bearing species. They are less abundant when carbon grains are destroyed. In Figure \ref{fig:11}, in the oxygen-rich case, the volatile water abundance is enhanced inside its snowline around 2 au. In the carbon-rich case, the gas-phase water abundance is very low inside 10 au and there is no clear water ice snowline present. The distributions of \ce{CO2} gas show clearly the snowline around 5 au in both cases. In the oxygen-rich case, the abundance peak of $\sim 10^{-4}$ appears near the midplane inside 2 au, whereas in the carbon-rich case, the peak value can only reach $\sim$$10^{-6}$ close to the central star. We note that the abundance of \ce{O2} gas can be affected by the initial abundance setting, atomic or molecular (\citealt{2016A&A...595A..83E}; \citealt{2018arXiv180803329E})

Finally, we briefly summarize in which molecules the carbon is mainly incorporated in both cases.
Excluding CO gas, in the oxygen-rich case, carbon mainly resides in \ce{CO2} gas inside the HCN snowline, and as \ce{CO2} gas and HCN ice beyond the HCN snowline. Beyond the \ce{CO2} snowline, carbon mainly resides in \ce{CO2} and HCN ice. A fraction of carbon forms \ce{CH4} gas throughout the inner disk. Meanwhile, in the carbon-rich case, carbon mainly forms HCN gas and long, gas-phase, carbon-chain molecules inside their snowlines. Beyond it, carbon mainly ends up in long, carbon-chain ice and gaseous \ce{CH4}, while inside the \ce{CO2} snowline, some carbon is converted to gas-phase \ce{CO2}.

\subsection{Radial distribution of solid carbon fraction} \label{subsec:3_2}

To compare the observed carbon depletion gradient in the inner solar system (Figure \ref{fig:1}) with the modeled values, we calculate the carbon fraction in grains relative to the solar abundance of silicon as a function of the disk radius,

\begin{equation}\label{eq:eq19}
\mbox{Solid Carbon Relative to Solar Silicon}= \frac{\mbox{Refractory\ Carbon+Carbon\ in\ Ice\ Mantle}}{\mbox{Solar\ Abundance\ of\ Silicon}} 
\end{equation}

We assume that the solar abundance of silicon is $3.55 \times 10^{-5}$ with respect to hydrogen nuclei (\citealt{2009ARA&amp;A..47..481A}; \citealt{draine2010physics}). The value for the refractory carbon is assumed to be the difference between the solar abundance of carbon, $2.95 \times 10^{-4}$ with respect to hydrogen nuclei (\citealt{draine2010physics}), and the gas-phase carbon abundance in the local diffuse ISM, $7.30 \times 10^{-5}$ (Table \ref{tab:1}). In the model with carbon grain destruction, all the carbon has been released from the refractory material and thus there is no carbon left in this form. The released refractory carbon will be incorporated into species either in the gas-phase or in the ice mantle and the abundance of ice mantle carbon is obtained from our model calculation described in Section \ref{subsec:3_1}. Table \ref{tab:3} shows the ratio of the carbon in the solid phase relative to the solar silicon at each disk radius.

\begin{table}[h]
\centering
\begin{tabular}{|c||c|c|c| }
\hline
 &  without destruction  & with destruction  &  with des. and increase $E_{bin}$ by 25\% \\
\hline 
10 au  &  6.25  & 3.62 & 3.69 \\ 
5 au   &  6.25 & 3.39  & 3.60\\
4 au   &  6.28  & 3.39 & 3.57 \\
3 au   &  6.28  & 3.38 & 3.56 \\
2 au   &  6.25  & 3.31  & 3.38 \\
1 au   &  6.25  & 0.04  &   $0.0015$\\
\hline 
\end{tabular}
\caption{The ratio of the carbon in the solid phase relative to the solar silicon at different disk radius.}
\label{tab:3}
\end{table}


\begin{table}[h]
\centering
\begin{tabular}{|c||c|c| }
\hline
& without destruction & with destruction \\
\hline
10 au  &  CO$_2$ HCN C$_5$         & C$_6$H C$_6$H$_2$ C$_2$H$_2$  \\ 
5 au   &  HCN C$_5$ HCOOH         & C$_6$H C$_6$H$_2$ C$_5$H  \\
4 au   &  HCN HCOOH  C$_7$         & C$_6$H C$_6$H$_2$ C$_6$  \\
3 au   &  HCN HCOOH  CO$_2$       & C$_6$H C$_6$H$_2$ C$_6$    \\
2 au   &  HCOOH  C$_9$  C$_2$S     & C$_9$H C$_9$H$_2$ C$_9$   \\
1 au   &  HCOOH  C$_9$H$_2$ CO$_2$ & C$_9$  C$_9$H$_2$ C$_1$$_0$   \\
 \hline 
\end{tabular}
\caption{The most abundant carbon-bearing species in the ice mantle in the midplane at different disk radii.}
\label{tab:4}
\end{table}

In the case without carbon grain destruction, the fraction does not change very much at different radii (Table \ref{tab:3}), since it is imposed that the majority of the carbon is locked in refractory form. Figure \ref{fig:12} shows the percentages of the form of carbon for the models with and without carbon grain destruction at a disk radius of 3 au as an example. It shows that most of the carbon ($\sim$75 \%) is locked in refractory form in the case without carbon grain destruction. In this case the elemental abundance of oxygen is larger than that of carbon (C/O $<$ 1) in the gas-phase, and thus most of the remaining carbon ($\sim$24\%) is stored in the form of CO gas with less than 1\% in the form of larger organic species. Since the refractory carbon and CO gas do not change significantly in abundance over 1 au $<$ r $<$ 10 au in this model, there is almost no fractional variation in this case. On the other hand, in the case with carbon grain destruction, the fraction varies slightly at 2 au $<$ r $<$ 10 au and suddenly drops at r = 1 au. In this case, all the carbon in refractory form released to the gas phase reacts to form a range of species. Since the elemental abundance of carbon is larger than that of oxygen in gas-phase (C/O $>$ 1), oxygen is mainly stored in CO in this case, capturing $1.76 \times 10^{-4}$ (see Table \ref{tab:1}), the corresponding amount of carbon, that is, $1.76 \times 10^{-4}/2.95 \times 10^{-4}\simeq$ 60 \% of the total carbon as CO gas (Figure \ref{fig:12}). The remaining carbon is mainly in the form of carbon-chain molecules in the ice mantle at $r\geq$ 2au, as is shown in Sect. \ref{subsec:3_1} (see e.g., Figure \ref{fig:9} and Figure \ref{fig:10}). Table \ref{tab:4} shows the most abundant carbon-bearing species in the ice mantle in the disk midplane at different radii. The carbon-bearing molecules in the ice mantle evaporate into the gas phase inside their snowlines and the locations of the snowlines depend on their binding energies (see Table \ref{tab:2}). Therefore, the fraction of carbon in the solid-phase changes across these snowlines of carbon-bearing species. Since the most abundant carbon-bearing molecules in the ice mantle are carbon-chain molecules (see Table \ref{tab:4}) whose snowlines are located around the disk radii of $\sim$ 2 au (see Table \ref{tab:2}), the solid carbon fraction does not decrease until r $<$ 2 au. Noted here, as we mentioned before, there are some uncertainties in the abundances of carbon-chain molecules in the ice mantle because grain surface reactions are not included in our calculation. Once grain surface reactions are considered, hydrogenation reactions on the surface may lead to the formation of alkanes.

While the model results reproduce qualitatively the trend of carbon depletion observed in the solar system, the comparison between our result (Table \ref{tab:3}) and the solar system data (Figure \ref{fig:1}) shows some quantitative discrepancies. There is a relatively high carbon fraction in the asteroid belt (a few to an order of magnitude larger) and too much depletion at 1 au ($\sim$ three orders of magnitudes smaller) compared to the observation values.
This is possibly caused by the omission of, for example, grain surface reactions and/or turbulent mixing, which we have not yet taken into account in our chemical model and will alter the chemical structure of the disk and the partitioning of carbon between solid and gaseous forms. With grain surface reactions, a higher carbon fraction in the ice mantle might be produced by large carbon-bearing molecules formed from species stuck on the grain surface that react with other species (e.g., \citealt{1992ApJS...82..167H}; \citealt{1993MNRAS.263..589H}; \citealt{2006A&amp;A...457..927G}; \citealt{2008ApJ...682..283G}; \citealt{Walsh_2014}). On the other hand, turbulent mixing can bring the dust grains from the midplane up to the warm surface layer where the ice mantle can be photodissociated or thermally desorbed. 
This leads to a decrease in the solid carbon fraction, perhaps to a level consistent with that in the asteroid belt (e.g., \citealt{2013ApJ...779...11F}; \citealt{2014ApJ...790...97F}; \citealt{2011ApJS..196...25S}). We note that the timescale of vertical mixing is $\tau_{mix}\sim 6\times 10^4 (r/10AU)(\alpha/0.01)^{-1}$ yr, where $\alpha$ represents the parameter for vertical mixing, similar to the $\alpha$ parameter in the viscous disk (e.g., \citealt{2013ApJ...779...11F}), and that the timescale is shorter in the inner region. In addition, uncertainties in the binding energies of molecules on grains will affect the solid carbon fraction in grains. Desorption rates depend exponentially on binding energies (see Equation \ref{eq:a2}).  However, due to the lack of appropriate data for many binding energies, the values are still very uncertain. We therefore enlarged the value of binding energies by 25\% to investigate how the solid carbon fraction changes as binding energies change. We found that the value at 1 au becomes comparable with the observed value on the Earth when binding energies increase about 25\% (see Table \ref{tab:3}). We note that the increased binding energies of carbon-chain molecules are close to the data in complied for the UMIST Database for Astrochemistry \footnote{\url{http://www.udfa.net/}} (\citealt{2013A&A...550A..36M}).

\subsection{Prediction for ALMA observations} \label{subsec:3_3}

Our results in Section \ref{subsec:3_1} suggest that if carbon grains are destroyed in the inner region of protoplanetary disks, it will affect the molecular abundance profiles, a process which could be tested by observing the molecular line emission. In this subsection, we perform ray tracing calculations, using the molecular abundance profiles obtained from our model calculation and determine whether or not carbon grain destruction in the inner disk can be tested observationally. 

While molecular lines have been observed towards protoplanetary disks at infrared and (sub)millimeter wavelengths, we choose lines which are observable with ALMA. Since, as we have seen in Section \ref{subsec:3_1}, significant differences in molecular abundances between the models with and without carbon grain destruction appear only near the disk midplane where CO is dominant and not photodissociated. Therefore, (sub)millimeter lines are more suitable to trace potential differences because infrared lines are optically thick and trace only the disk surface layer. In addition, ALMA enables us to spatially resolve the inner region of the disk, which allows us to see the difference between the models more clearly. Among the molecular lines observable at (sub)millimeter wavelengths, we choose the HCN lines because (1) the HCN abundance distribution is very much affected by the carbon grain destruction (see Section \ref{subsec:3_1}), and (2) they are known to be strong towards protoplanetary disks (e.g., \citealt{1997A&amp;A...317L..55D}; \citealt{2008ApJ...681.1396Q}; \citealt{2010ApJ...720..480O}; \citealt{2011ApJ...734...98O}; \citealt{2012A&amp;A...537A..60C}; \citealt{2017ApJ...835..231H}).
Three HCN isotopologues, DCN, \ce{H^{13}CN} and \ce{HC^{15}N}, have been detected towards protoplanetary disks (e.g., \citealt{2008ApJ...681.1396Q}; \citealt{2017ApJ...835..231H}; \citealt{2015ApJ...814...53G}; \citealt{2017ApJ...836...30G}). We concentrate here on \ce{H^{13}CN} because the DCN/HCN ratio is known to be sensitive to the temperature profile (e.g., \citealt{2001A&amp;A...371.1107A}; \citealt{2007ApJ...660..441W}; \citealt{2016ApJ...819...13C}; \citealt{Aikawa_2018}), while \ce{H^{13}CN} is less affected (e.g., \citealt{2009ApJ...693.1360W}). Meanwhile, \ce{H^{13}CN} lines are stronger than \ce{HC^{15}N} lines (\citealt{2017ApJ...836...30G}) and it is easier to analyze its spatial distribution. The combination of HCN and its isotopologue lines is useful to test the carbon grain destruction model since HCN lines are optically thick and trace mostly surface layer of the disk where CO is photodissociated and the carbon grain destruction does not affect molecular abundances significantly. The HCN isotopologue lines are less optically thick and trace the lower layer of the disk where CO is not photodissociated and the effect of carbon grain destruction is significant. 

Figure \ref{fig:13} displays the profiles of the HCN and \ce{H^{13}CN} J=4-3 lines at 354.5 and 345.3 GHz respectively, with the spectral resolution of 0.4 $\mbox{km/s}^{-1}$, for the T-Tauri disk, calculated by performing ray tracing calculations (see Section \ref{subsec:3_2}) using the obtained physical structure (Figure \ref{fig:2}) and the HCN abundance profile (Figure \ref{fig:7}). The HCN/\ce{H^{13}CN} abundance ratio is simply assumed to be the typical interstellar value of 70 (\citealt{2011ApJ...740...84Q}). According to \cite{2009ApJ...693.1360W}, it could be larger by a factor of $\leq$ 2 in the disk surface where the effect of self-shielding is significant and different CO isotopologues are selectively photo-dissociated. But the assumption is reasonable in the region closer to the disk midplane. We assume the distance to the T-Tauri disk is 70 pc and its inclination angle is 30 degrees. The left- and right-hand sides of Figure \ref{fig:13} show the line profiles of HCN and \ce{H^{13}CN}, respectively. The line profiles are calculated only inside a radius of 5 au since the significant differences between two models appears mainly inside 2 au (Figure \ref{fig:7}). Solid and dashed lines represent models with and without grain destruction, respectively. Figure \ref{fig:13} indicates that the line emission of \ce{H^{13}CN} shows more obvious differences between the two cases. However, the peak flux density is less than 1 mJy for the \ce{H^{13}CN}  line for the model without the carbon grain destruction, and it is too weak to use it for testing the carbon grain destruction model even with ALMA.

For this reason, we have also modeled a Herbig Ae disk since the radiation from the central source is stronger and the hot region is larger (Figure \ref{fig:2}). Thus, the carbon grain destruction region spreads to larger radii and the observational test will be easier. Figure \ref{fig:14} shows the distribution of HCN in the Herbig Ae disk as a function of radius up to 50 au and the disk height divided by the radius (Z/r). Gas-phase reactions produce HCN efficiently only in the very hot region inside a disk radius of r $\leq$ 5 au for the model without carbon grain destruction, while gas-phase HCN can be very abundant in the whole region inside the HCN snowline at $\sim$ 20 au for the model with carbon grain destruction. Therefore, a dramatic change appears appears at radii of 2 - 30 au between two models. In contrast, the significant differences in T-Tauri disk appear only very close to the central star (inside 2 au) and the flux density is too low to be observed. 

Figure  \ref{fig:15} is the zeroth moment map, that is, the integrated intensity map of the HCN and \ce{H^{13}CN} lines, calculated as 
\begin{equation}\label{eq:eq20}
M_0(x,y)=\int I(x,y,v)dv
\end{equation}
where $I(x,y,v)$ is the intensity as a function of position, $(x,y)$, and velocity, $v$, given by Equation \ref{eq:eq15}.

The physical structure (Figure \ref{fig:2}) and the HCN abundance profile (Figure \ref{fig:14}) of the Herbig Ae disk are used. We assume that the distance to the disk is 140 pc and its inclination angle is 30 degrees. The \ce{H^{13}CN} line intensity map shows the difference more clearly than the HCN line. This is because the \ce{H^{13}CN} line traces a relatively lower layer of the disk due to relatively lower optical depth than the HCN line. Figure \ref{fig:16} shows the optical depth of the HCN and \ce{H^{13}CN} lines along the vertical direction of the disk for the models without and with carbon grain destruction. The HCN line becomes optically thick up to a disk radius of $\sim$ 30 au even for the model without the carbon grain destruction, that is, the HCN line only traces the disk surface where the difference in the HCN abundance is not significant between the models (Section \ref{subsec:3_1}). On the other hand, the isotopologue \ce{H^{13}CN} line is optically less thick and can trace the HCN abundance difference near the midplane at a disk radii of r $<$ 30 au. Therefore, the difference in the emitting regions of intensity map between the HCN and \ce{H^{13}CN} lines is a good tracer for testing the carbon grain destruction. 

Even in those cases for which spatially resolved observations are difficult, the profiles of the HCN and \ce{H^{13}CN}  lines can be a good tracer as well. Figure \ref{fig:17} shows the line profiles of the HCN line (left-hand-side) and the \ce{H^{13}CN} line (right-hand-side). In both profiles, the solid line represents the model with carbon grain destruction and the dotted line is the model without. Like the T-Tauri disk model (Figure \ref{fig:13}), the line emission of \ce{H^{13}CN} shows more substantial differences between the two models. The differences in the \ce{H^{13}CN} line range from high velocity ($\sim$10 km/s, tracing the inner region) to low velocity (tracing the outer region). In contrast, the differences in HCN emission mainly exist at velocity of -5 to + 5 km/s.

Figure \ref{fig:18} is the normalized cumulative line flux as functions of disk radius (left-hand-side) and the velocity (right-hand-side). The former is calculated from the disk center to the disk radius of 25 au and the latter is calculated over the range 0 $<$ $v$ $<$ 20 km/s. Each cumulative flux is normalized using the following equations,
\begin{equation}\label{eq:er21}
\begin{split}
F_{cum}(r)=\frac { \int _{ 0 }^{ r } 2\pi r'dr' \int_{-a}^{+a} I(r',v)dv  }
{ \int _{ 0 }^{ b} 2\pi r'dr' \int_{-a}^{+a} I(r',v)dv  }, \\
F_{cum}(v)=\frac { \int _{ 0 }^{ v } dv' \int _{ 0 }^{ b} I(r,v') 2\pi rdr } 
{ \int _{ 0 }^{ +a} dv' \int _{ 0 }^{b}  I(r,v') 2\pi rdr },
\end{split}
\end{equation}
where $a$ is defined as 20 km $\mbox s^{-1}$ and $b$ as 25 au. The blue and green lines are models with and without carbon grain destruction, respectively, with HCN represented by the solid lines and \ce{H^{13}CN} the dashed lines. Comparing the results in the same color, the differences between two models can be easily distinguished by the ratio of HCN and \ce{H^{13}CN}. In left-hand-side figure, the green dashed line reaches $\sim$ 60 \% of the cumulative flux inside 5 au and the green solid line reaches the same level inside $\sim$ 15 au. This can be explained from the intensity map. For the model without carbon grain destruction, the intensity of the \ce{H^{13}CN} line is strong only in very compact region ($r$ $<$ 5 au) as shown in Figure \ref{fig:15}. Meanwhile, the HCN line is strong out to $r$ $\sim$ 30 au and thus the normalized cumulative flux increases smoothly with radius. On the other hand, for the model with carbon grain destruction, both HCN and \ce{H^{13}CN} lines are strong out to $r$ $\sim$ 30 au and the cumulative fluxes increase smoothly, with no significant difference between them, so that differences between the two models can be tested by the cumulative flux as a function of radius if it can be spatially resolved. Even in the case for which the emission is not spatially resolved, we can see these differences in the line profiles, that is, cumulative flux as a function of velocity (e.g., \citealt{Zhang_2017}). The green lines in Figure \ref{fig:18} (right-hand-side) show the differences between two models clearly. Comparing the normalized cumulative flux with the line profiles (Figure \ref{fig:17}), the shape and the distribution of the model without carbon grain destruction are very different from HCN and \ce{H^{13}CN}. However, the line shape and the distribution in the model with carbon grain destruction are very similar for both HCN and \ce{H^{13}CN}. Both figures show significant differences between the HCN and \ce{H^{13}CN} lines for the model without the carbon grain destruction, while the difference is not large for the model with carbon grain destruction. Therefore, the differences between two models can be easily distinguished by the ratio of the HCN and \ce{H^{13}CN} line.

HCN and \ce{H^{13}CN} lines have been detected toward some Herbig Ae stars by ALMA (e.g., \citealt{2015ApJ...814...53G}, \citealt{2017ApJ...835..231H}). \citet{2015ApJ...814...53G} have reported the observations of the HCN J = 4 - 3 and \ce{H^{13}CN} J = 3 - 2 lines with relatively low spatial resolution of $>$ 0.7\arcsec. Mapping the inner region of the disk (around the disk radius of 10 - 20 au), taking advantage of ALMA's high spatial resolution and high sensitivity, would be useful to diagnose carbon grain destruction in the inner disk.

Because the abundance distribution of molecules has some uncertainty depending on the chemical model, here we suggest c-\ce{C_3H_2} $6_{1,6}$-$5_{0,5}$ (217.882 GHz) as the other target line. This line has been detected in a Herbig Ae disk (\citealt{2013ApJ...765L..14Q}). The line is blended with the c-\ce{C3H2} $6_{0,6}$-$5_{1,5}$ line, but we treat only the \ce{C_3H_2} $6_{1,6}$-$5_{0,5}$ since it is slightly stronger. We note that other transition lines of c-\ce{C3H2} have been detected toward T Tauri disks (\citealt{Bergin_2016}). 
Figure \ref{fig:19} shows that the distribution of c-\ce{C_3H_2} in the Herbig Ae disk as a function of radius up to 50 au and the disk height divided by the radius (Z/r), which indicates the significant difference appears at r $<$ 30 au between the models, similar to the case of HCN. The difference in the abundance distribution results in the clear differences in both the intensity map (Figure \ref{fig:20}) and the line profile (Figure \ref{fig:21}). To sum up briefly, we suggest \ce{HCN}, \ce{H^13CN} and c-\ce{C_3H_2} as possible tracers of testing the carbon grain destruction effect in protoplanetary disks.

\section{Summary} \label{sec:4}

In this work, we focus on two issues: (1) The carbon depletion gradient in the inner solar system, and (2) searching for observational evidence of carbon grain destruction in the disks of T-Tauri and Herbig Ae stars. We conclude that the carbon grain destruction affects the abundances and distribution of various molecules in the protoplanetary disk, showing significant differences especially near the midplane in the inner region of the disk, where CO gas is abundant and is not photodissociated. For example, the gas-phase HCN abundance shows 8 orders of magnitudes difference near the midplane inside the radius of 2 au in the T-Tauri disk. 

The distribution of molecules is determined by their volatility such that volatile species can evaporate into gas even in the outer region of the disk. Those molecules which remain in ice mantles can influence the composition of subsequently forming planets. Therefore, we present the location of the snowlines and calculate the solid carbon fraction relative to the solar abundance of carbon as a function of the radius in the inner region of a protoplanetary disk. Although the carbon depletion gradient is reproduced in the model with carbon grain destruction, the resulting solid carbon fractions show a quantitative discrepancy from those in the solar system. The solid carbon fractions in the asteroid belt are about an order of magnitude larger than the measured value of the meteorites. Meanwhile, the value at 1 au is 3 orders of magnitude smaller than the measured value. Including grain surface reactions in the model may help to better reproduce the carbon depletion gradients. For example, grain surface reactions are expected to make more complex, less volatile, organic molecules in the disk and could further enlarge the solid carbon fraction at 1 au. 
We also examined the effect of carbon grain destruction on predictions for ALMA observations. We find that lines in T Tauri are too weak to probe carbon grain destruction but that ALMA can probe this effect through the line ratio of HCN/\ce{H^{13}CN} as well as c-\ce{C3H2} in Herbig Ae disks.


\acknowledgments

We thank an anonymous referee for comments which improved the content of the manuscript. We are grateful to Patrice Theule and Sheng-Yuan Liu for discussions in Theoretical Institute for Advanced Research in Astrophysics (TIARA) and Academia Sinica Institute of Astronomy and Astrophysics (ASIAA). We acknowledge Yukawa Institute for Theoretical Physics (YITP) in Kyoto University and the Center of Computational Astrophysics (CfCA) at National Astronomical Observatory of Japan (NAOJ) for carrying out the numerical calculations. This work is supported by JSPS/MEXT Grants-in-Aid for Scientific Research 25108004, 25400229, and 15H03646 and Astrophysics at Queen's University Belfast is supported by a grant from the STFC (ST/P000321/1). JEL is supported by the Basic Science Research Program through the National Research Foundation of Korea (grant No. NRF-2018R1A2B6003423) and the Korea Astronomy and Space Science Institute under the R\&D program supervised by the Ministry of Science, ICT and Future Planning. CW acknowledges financial support from the University of Leeds and STFC (grant number ST/R000549/1).

\appendix

\section{The calculation of  adsorption and desorption rates } \label{appen_a}

The adsorption(freeze-out) rate of species i onto grain surface, $k_i^a  \mbox{ [s}^{-1}\mbox{]}$, is written as (\citealt{1992ApJS...82..167H})

\begin{equation} \label{eq:a1}
k_i^a=\alpha \sigma_d<v_i^{th}>n_d \quad s^{-1},
\end{equation}
where $\alpha$ is the sticking coefficient and we set it as 0.4 for all species (\citealt{2014ApJ...790....4V}), $\sigma_d=\pi r_d^2$ is the geometrical cross section of a dust grain, $r_d$ is the dust grain radius and $n_d$ is the number density of dust grains. We fix the number density ratio of dust grains to hydrogen nuclei ($d_g=n_d/n_H$) times the geometrical cross section of grain ($\sigma_d$), as $\pi<d_gr_d^2>=6.9 \times 10^{-22}$ $\mbox{[cm}^2\mbox{]}$, according to \citet{1992MNRAS.255..471R}, which is consistent with a gas-to-dust mass ratio of 100. $<v_i^{th}> = (k_BT_g/m_i)^{1/2}$ represents the thermal velocity of species $i$, where $k_B$ is the Boltzmann's constant, $T_g$ is the temperature of gas and $m_i$ is the mass of species $i$.

The thermal desorption rate, $k_i^d$ $\mbox{[s}^{-1}\mbox{]}$, with which species $i$ evaporate from the dust grain surface, is represented as (\citealt{1992ApJS...82..167H})

\begin{equation}\label{eq:a2}
k_i^d=\nu_0(i)\exp(\frac{-E_d(i)}{T_d}) \quad s^{-1},
\end{equation}
where $E_d(i)$ is the binding energy of species $i$ to the dust grain surface in units of Kelvin, $T_d$ is the dust temperature and $\nu_0$ is the vibrational frequency of each adsorbed species in its potential well (Hasegawa et al. 1992).

\begin{equation}\label{eq:a3}
\nu_0(i)=\sqrt{\frac{2n_sk_BE_d(i)}{\pi^2m_i}} \quad s^{-1}
\end{equation}
where $n_s$ represents the surface density of sites, $n_s=1.5\times10^{15}$ $\mbox{cm}^{-2}$, and $m_i$ is the mass of the adsorbed species $i$. 

For cosmic-ray induced thermal desorption, we assume that dust grains with a radius of 0.1 $\micron$ are heated by the impact of relativistic Fe nuclei with energy from 20 to 70 MeV $\mbox{nucleon}^{-1}$ and deposit an energy of 0.4 MeV on average into each dust grain (\citealt{1985A&amp;A...144..147L}; \citealt{1993MNRAS.263..589H}). Assuming that the majority of molecules will desorb around 70 K, the cosmic ray induced thermal desorption rate, $k_i^{crd}$, is expressed as 

\begin{equation}\label{eq:a4}
k_i^{crd} \approx f(70K)k_i^d(70K)\frac{\zeta_{CR}}{1.36\times10^{-17}s^{-1}} \quad s^{-1},
\end{equation}
where $\zeta_{CR}$ is the cosmic ray ionization rate of \ce{H2} scaled by the interstellar value used in the UMIST database, $1.36 \times 10^{-17}$ $\mbox{s}^{-1}$. $k_i^d \mbox{(70 K)}$ is the thermal desorption rate of species $i$ at dust temperature of 70 K. $f(70 K)$ is the fraction of time that dust grain spends above 70 K and it is roughly calculated by the ratio of the desorption cooling time ($\sim 10^{-5}$ \mbox{s}) to the total interval time for the temperature of dust grain to become 70 K ( $3.16 \times 10^{13}$ \mbox{s}, \citealt{1985A&amp;A...144..147L}). Therefore, f(70 K) $\thickapprox 3.16 \times 10^{-19}$. Although X-rays can penetrate the disk and induce desorption as well, we do not include it in this chemical network because of the remaining uncertainty in the desorption rate.  

Photodesorption is independent of the surface binding energy. The photodesorption rate adopted is based on the experiments of \citet{1995P&amp;SS...43.1311W} and \citet{2007ApJ...662L..23O}. Their results show that each photon absorbed by a dust grain will release a particular number of molecules into the gas phase and it is related to the fractional abundance on the dust grain surface. The photodesorption rate, $k_i^{ph}$, is expressed by the following equation (\citealt{2000ApJ...544..903W}; \citealt{2007ApJ...660..441W})  

\begin{equation}\label{eq:a5}
k_i^{ph} = F_{UV}Y_{UV}^i\sigma_d \frac{n_d}{n_{act}} \quad s^{-1},
\end{equation}
where $F_{UV}$ represents the UV radiative field at each position in the disk in units of photon $\mbox{cm}^{-2}\mbox{s}^{-1}$. $Y_{UV}^i$ is the photodesorption yield determined from the experiments in the unit of molecules $\mbox{photon}^{-1}$, and we adopt the value of $3.0 \times 10^{-3}$, which is determined by experiments of pure water ice (\citealt{1995P&amp;SS...43.1311W}) and pure CO ice (\citealt{2007ApJ...662L..23O}). $n_{act}$=$4 \pi r_d^2n_dn_sN_{LAY}$ represents the number of active surface sites in the ice mantle per unit volume and $N_{LAY}$ is the number of surface layers considered as active, assumed to be two.





\bibliographystyle{aasjournal}
\bibliography{manuscript_carbon}



\begin{figure} [b]
\centerline{\includegraphics[width=0.8\textwidth]{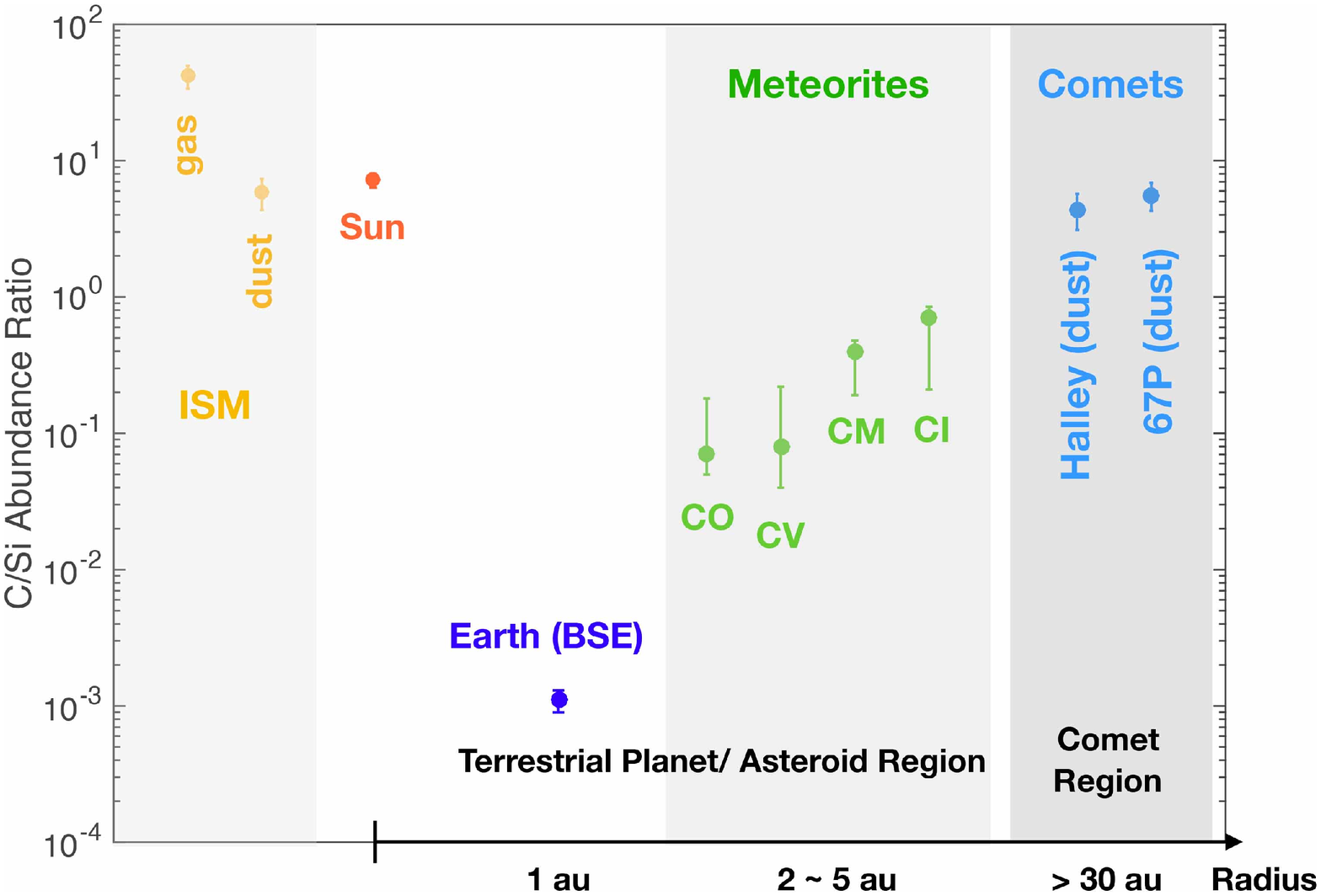}}
\caption{The carbon to silicon abundance ratio with error bars in the protosun (\citealt{2010ASSP...16..379L}), Earth, four classes of carbonaceous chondritic meteorites (CI, CO, CM and CV; \citealt{2015PNAS..112.8965B}), cometary dust of Halley (\citealt{1988Natur.332..691J}) and 67P/C-G (\citealt{2017MNRAS.469S.712B}), and ISM (\citealt{2015PNAS..112.8965B}). }
\label{fig:1}
\end{figure}

\begin{figure} [b]
\centerline{\includegraphics[width=0.7\textwidth]{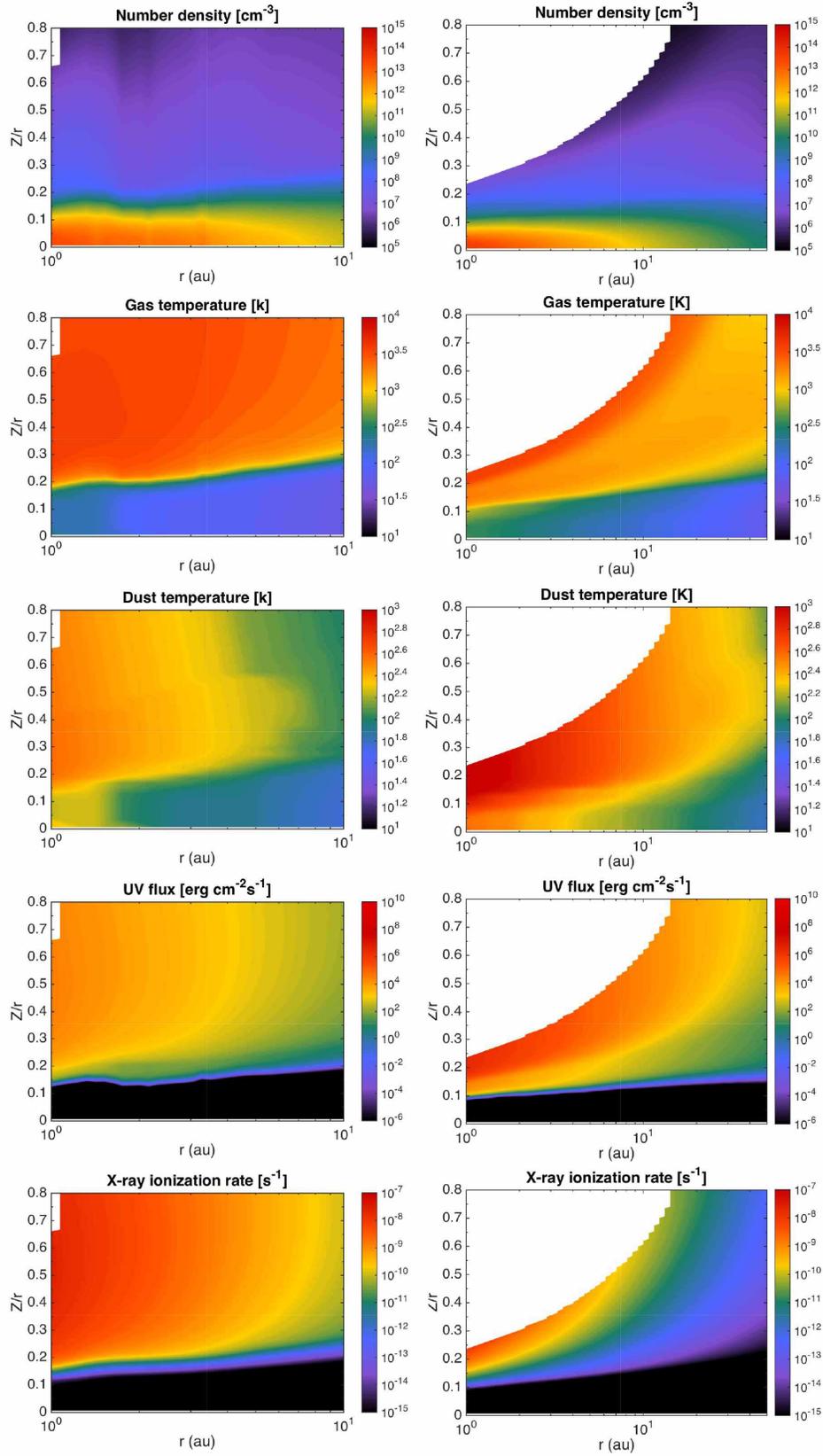}}
\caption{The gas number density, gas temperature, dust temperature, UV flux and X-ray ionization rate as a function of radius and height divided by the radius (Z/r) for the T Tauri disk (left-hand-side) and the Herbig Ae disk (right-hand-side). Noted that the radial ranges are different between the two disks.}
\label{fig:2}
\end{figure}

\begin{figure}
\centerline{\includegraphics[width=1 \textwidth]{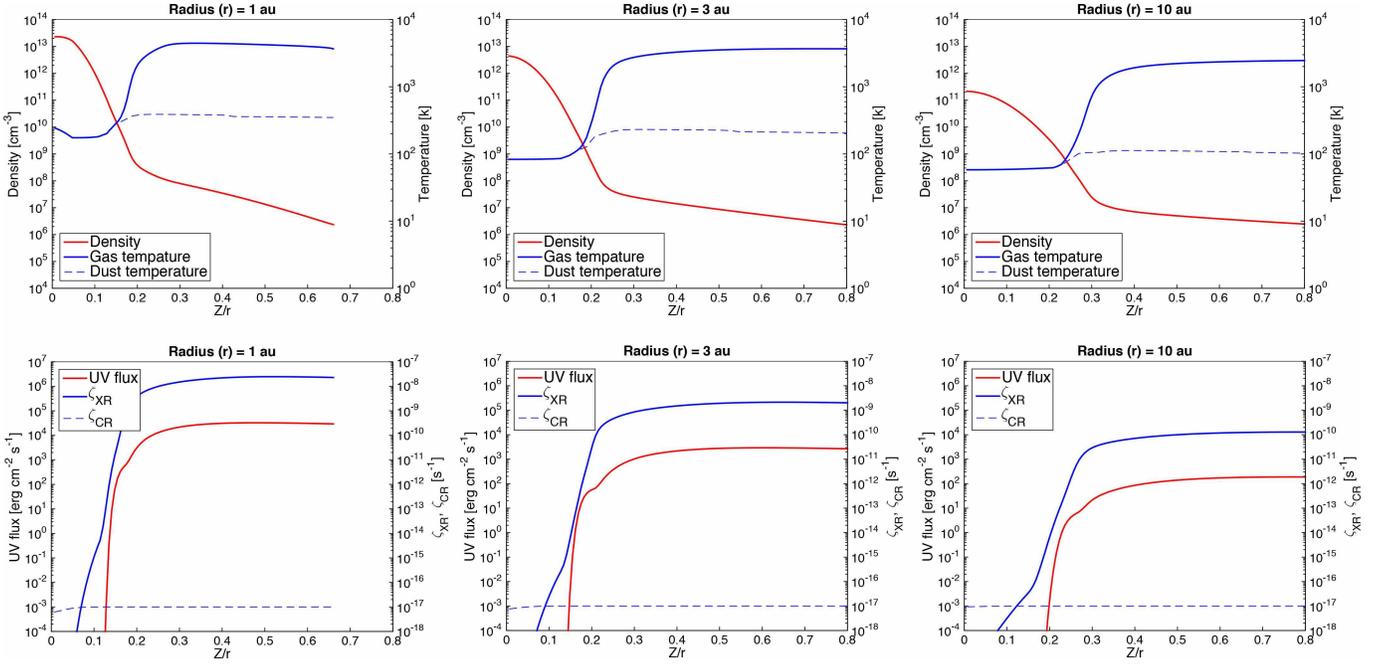}}
\caption{The top panels are number density of hydrogen nuclei (red line), the gas temperature (blue solid lines) and dust temperature (blue dashed lines) as a function of height divided by radius (Z/r) at a disk radius of 1 au (left), 3 au (middle) and 10 au (right). The bottom panels show the UV flux (red line), X-ray ionization rate (blue solid lines) and the cosmic-ray ionization rate (blue dashed lines) as a function of height divided by radius (Z/r) at a disk radius of 1 au (left), 3 au (middle) and 10 au (right) for a T Tauri disk.}
\label{fig:3}
\end{figure}

\begin{figure}
\centerline{\includegraphics[width=1 \textwidth]{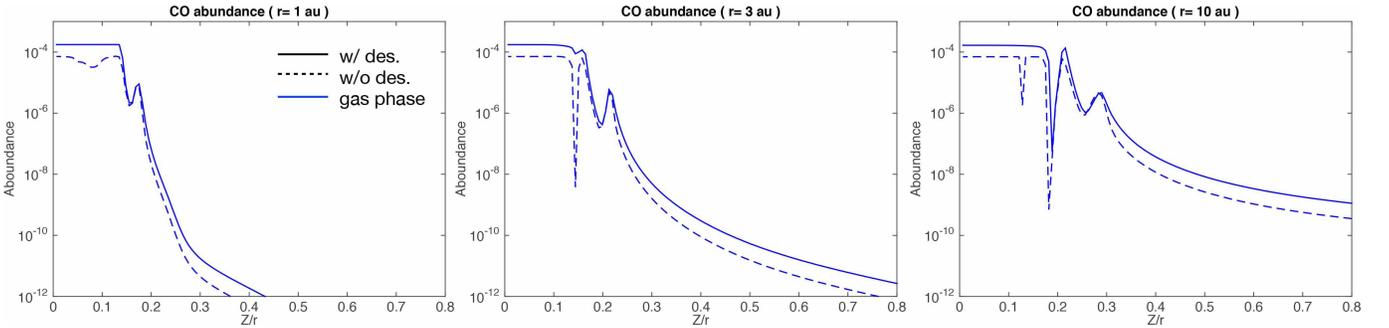}}
\caption{The fractional abundance of CO as a function of height divided by radius (Z/r) at a disk radius of 1 au (left), 3 au (middle) and 10 au (right). The blue lines show CO gas and light blue lines show CO in ice mantle for a T Tauri disk. Solid and dashed lines represent the model with and without carbon grain destruction, respectively.}
\label{fig:4}
\end{figure}

\begin{figure}
\centerline{\includegraphics[width=0.85 \textwidth]{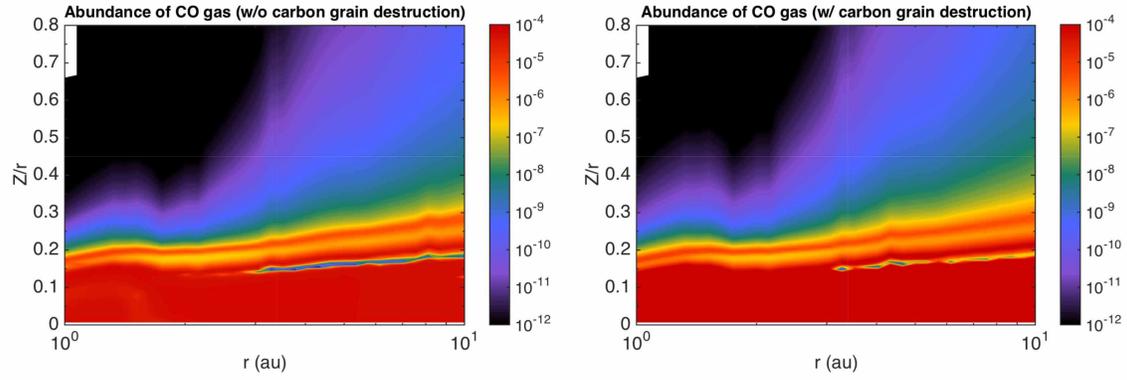}}
\caption{The 2-dimensional abundance distribution of gas-phase CO for the models without (left) and with (right) carbon grain destruction as a function of radius and height of the disk for a T Tauri disk. A detailed explanation of CO gas depletion layers can be found in Sec. \ref{subsec:3_1}.}
\label{fig:5}
\end{figure}

\begin{figure}
\centerline{\includegraphics[width=0.85 \textwidth]{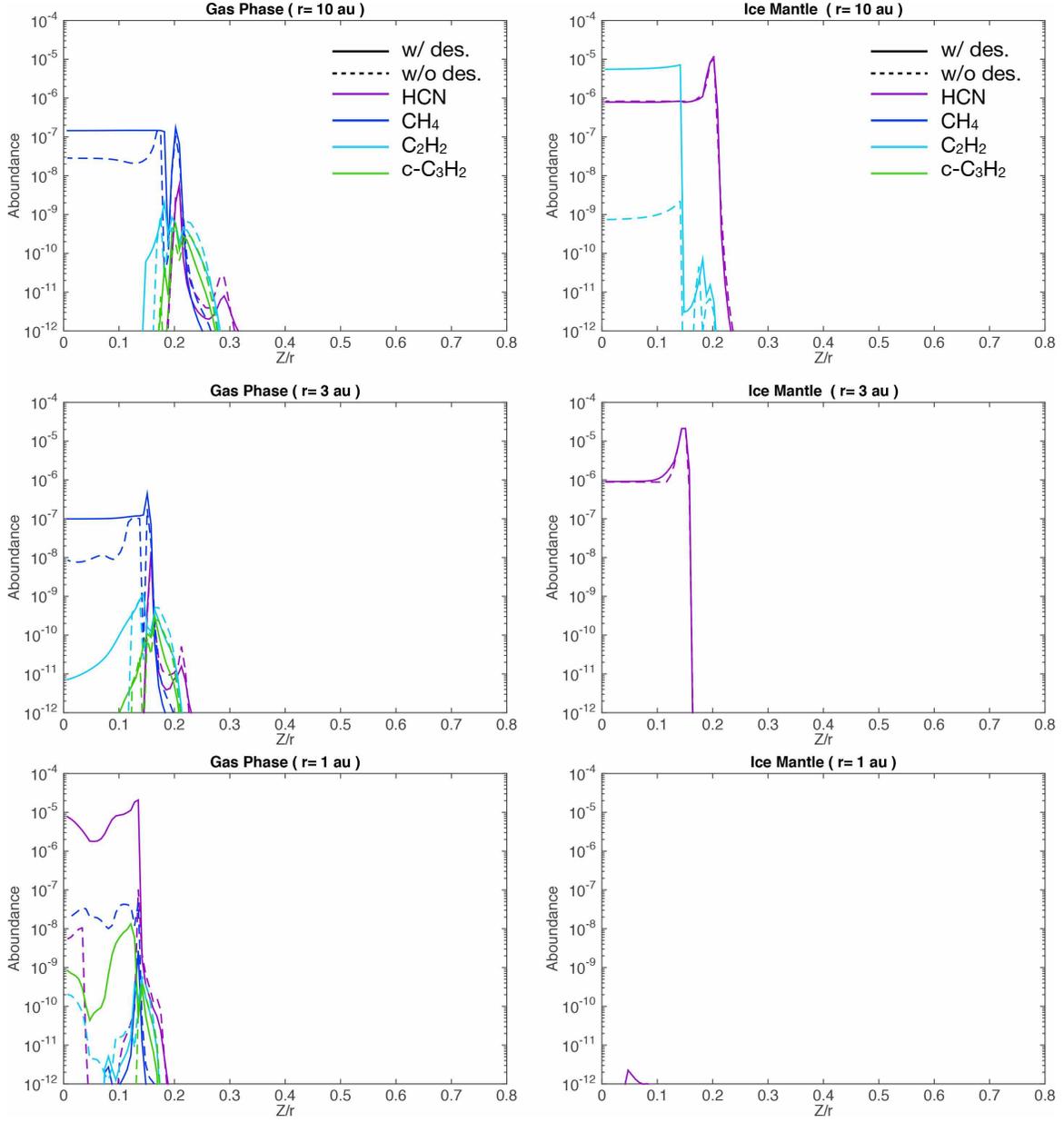}}
\caption{The abundances of carbon-bearing molecules as a function of height divided by radius (Z/r) at a disk radius of 10 au (top), 3 au (middle) and 1 au (bottom) for a T Tauri disk. The left and right panels show the molecular abundances in gas and ice, respectively. Solid and dashed lines represent the models with and without carbon grain destruction, respectively.}
\label{fig:6}
\end{figure}

\begin{figure}
\centerline{\includegraphics[width=0.85 \textwidth]{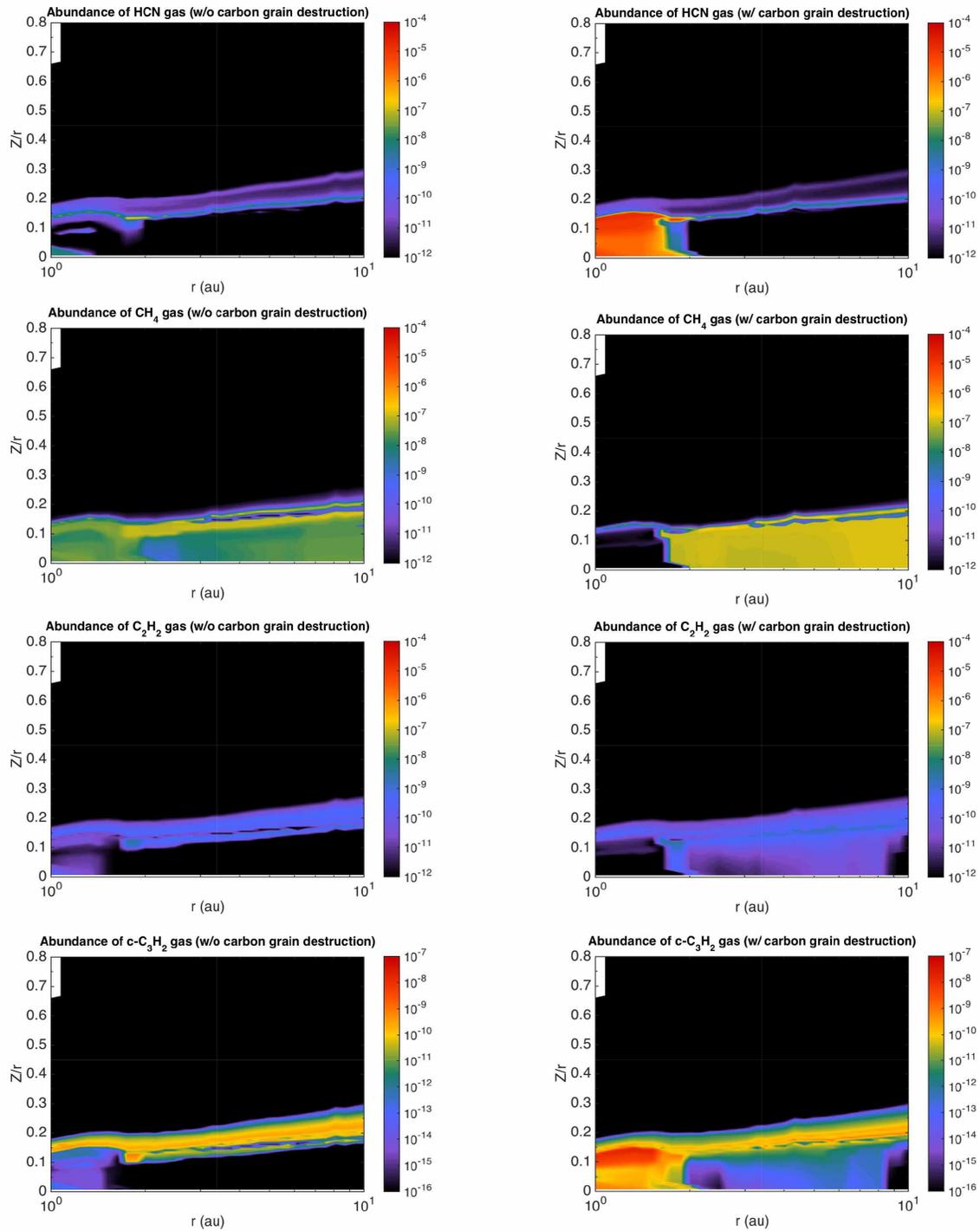}}
\caption{The 2-dimensional abundance distribution of gas-phase carbon-bearing molecules for the models without (left) and with (right) carbon grain destruction as a function of radius and height of the disk for a T Tauri disk. Noted that the colarbar range of c-\ce{C3H2} is different from the others ($10^{-16}$ to $10^{-7}$).}
\label{fig:7}
\end{figure}

\begin{figure}
\centerline{\includegraphics[width=0.85 \textwidth]{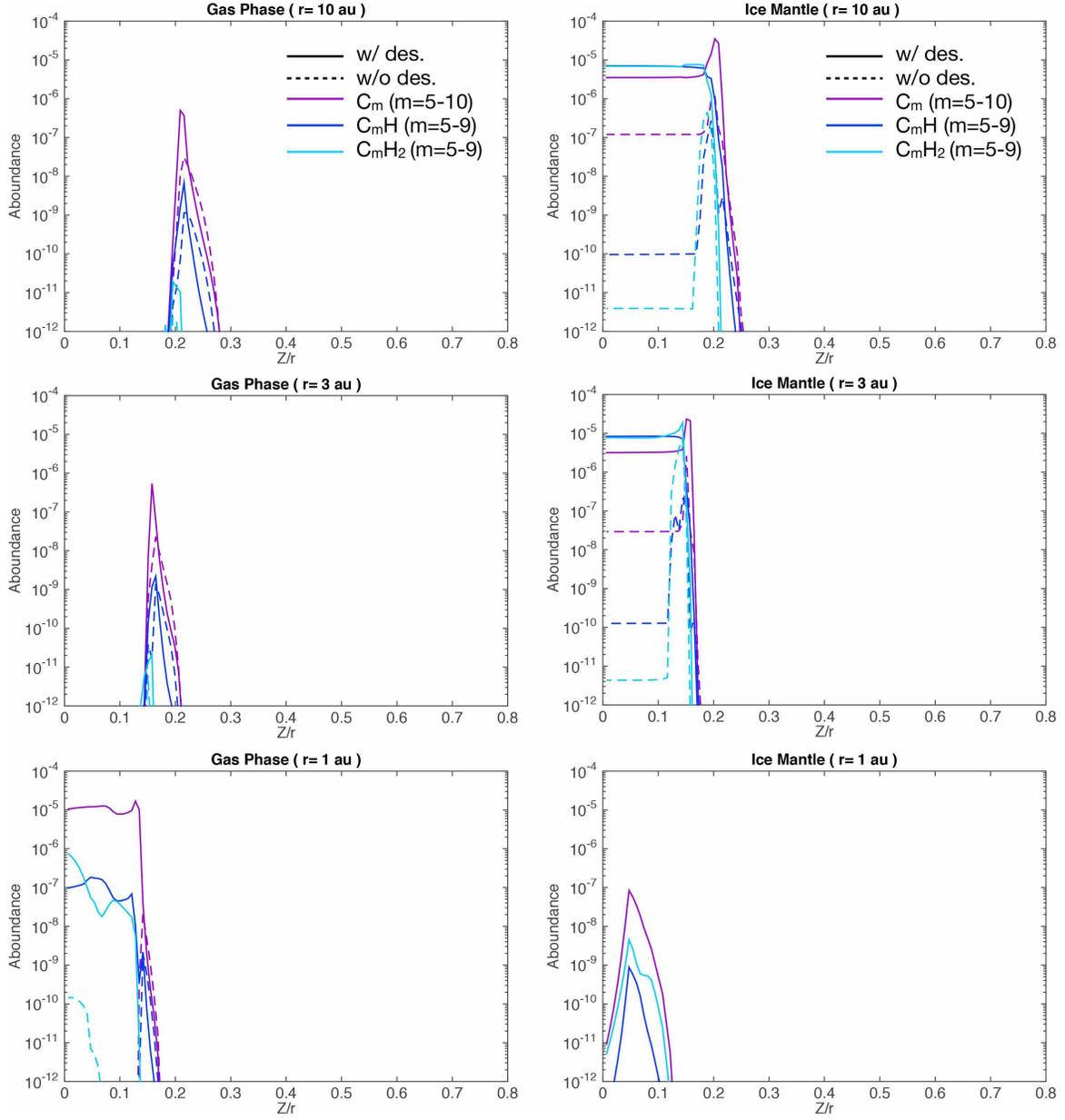}}
\caption{The same as Figure 6 but for the fractional abundance of \mbox{C$_m$}, \mbox{C$_m$H} and \mbox{C$_m$H$_2$}, summed over $m$ as indicated in the figure.}
\label{fig:8}
\end{figure}

\begin{figure}
\centerline{\includegraphics[width=0.85 \textwidth]{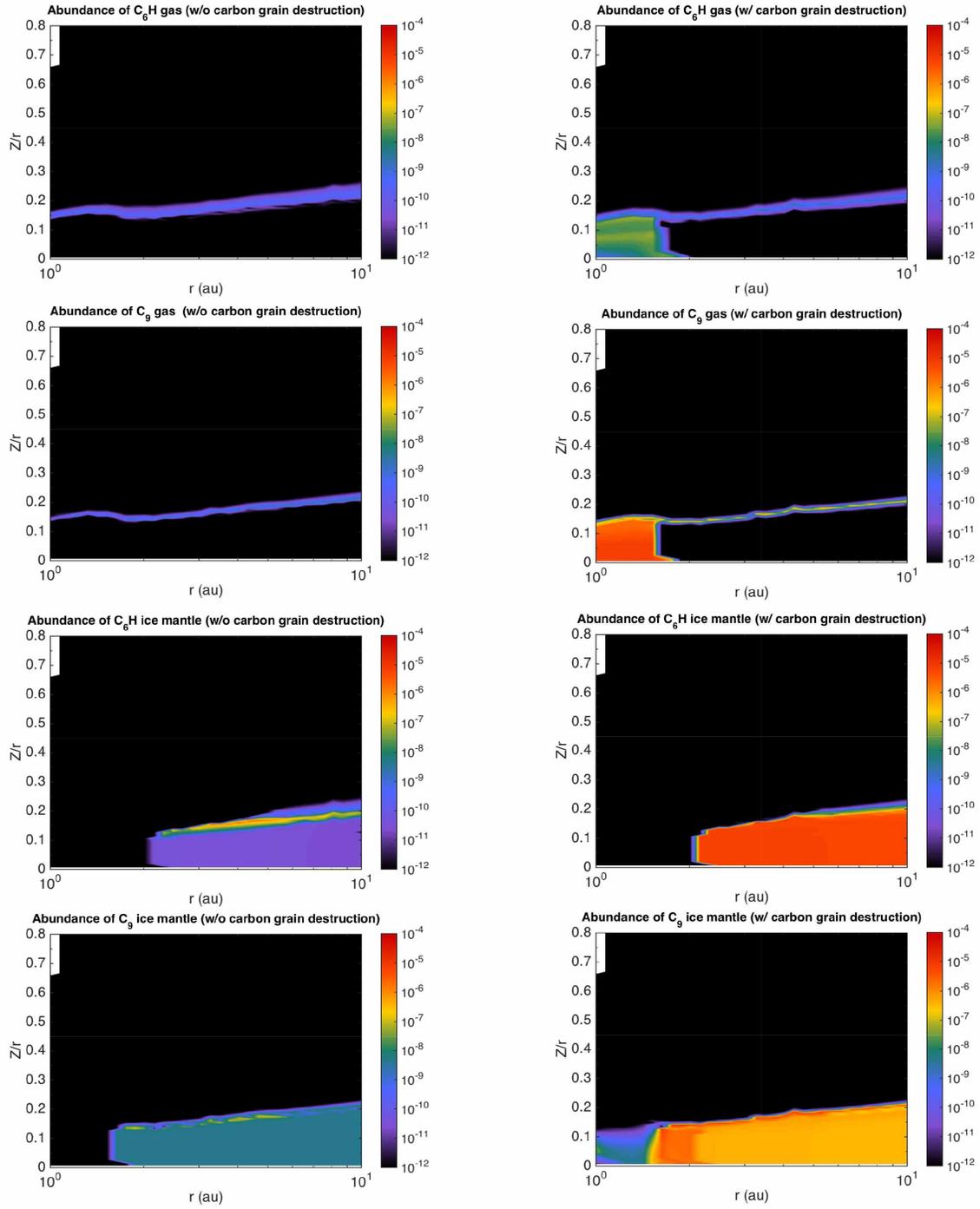}} 
\caption{The 2-dimensional abundance distribution of \ce{C6H} and \ce{C9} in the gas-phase and in the ice mantle for the models without (left) and with (right) carbon grain destruction as a function of radius and height of the disk for a T Tauri disk.}
\label{fig:9}
\end{figure}

\begin{figure} 
\centerline{\includegraphics[width=0.85\textwidth]{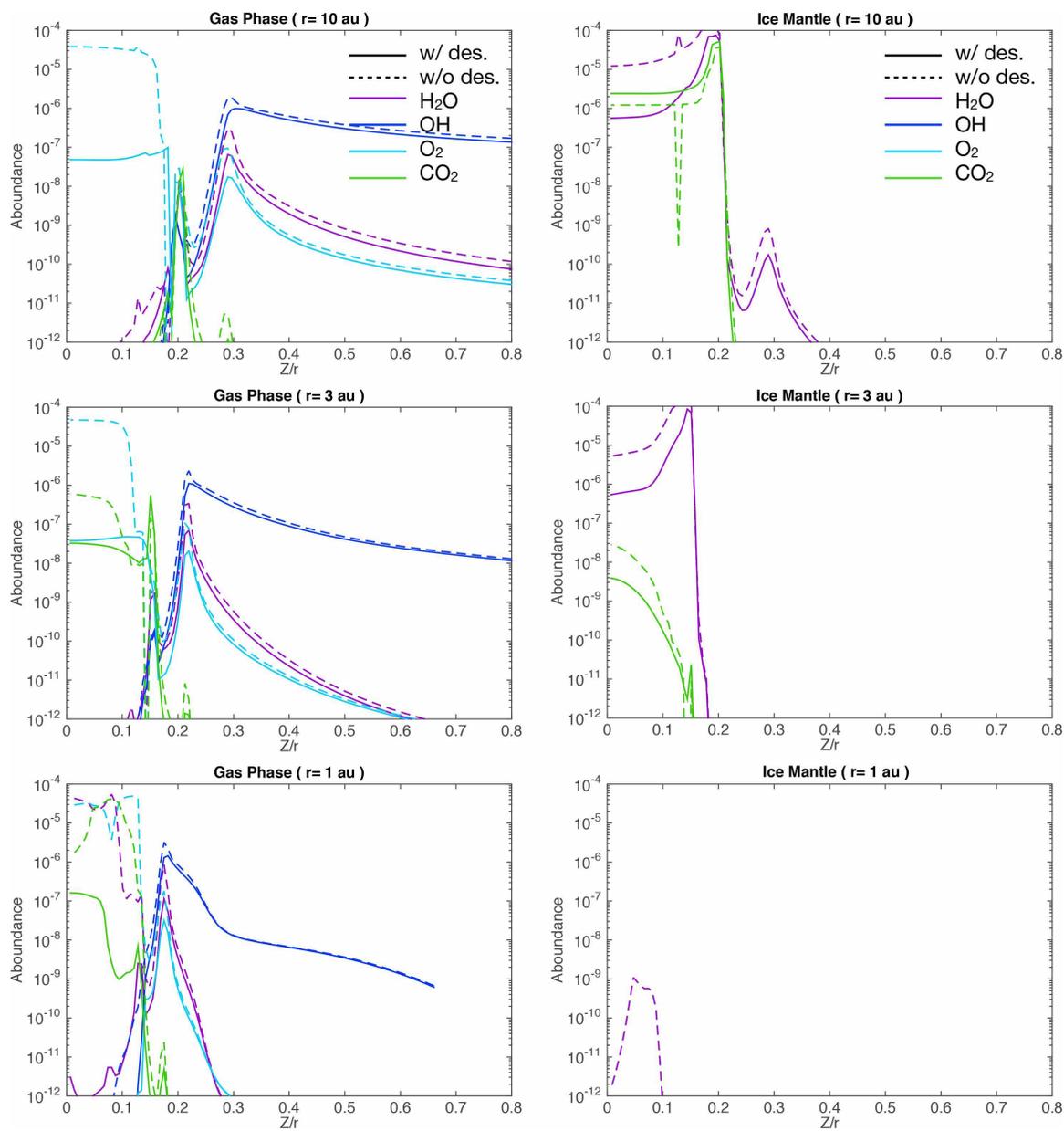}}
\caption{The same as Figure 6 but for the oxygen-bearing molecules.}
\label{fig:10}
\end{figure}

\begin{figure} 
\centerline{\includegraphics[width=0.85\textwidth]{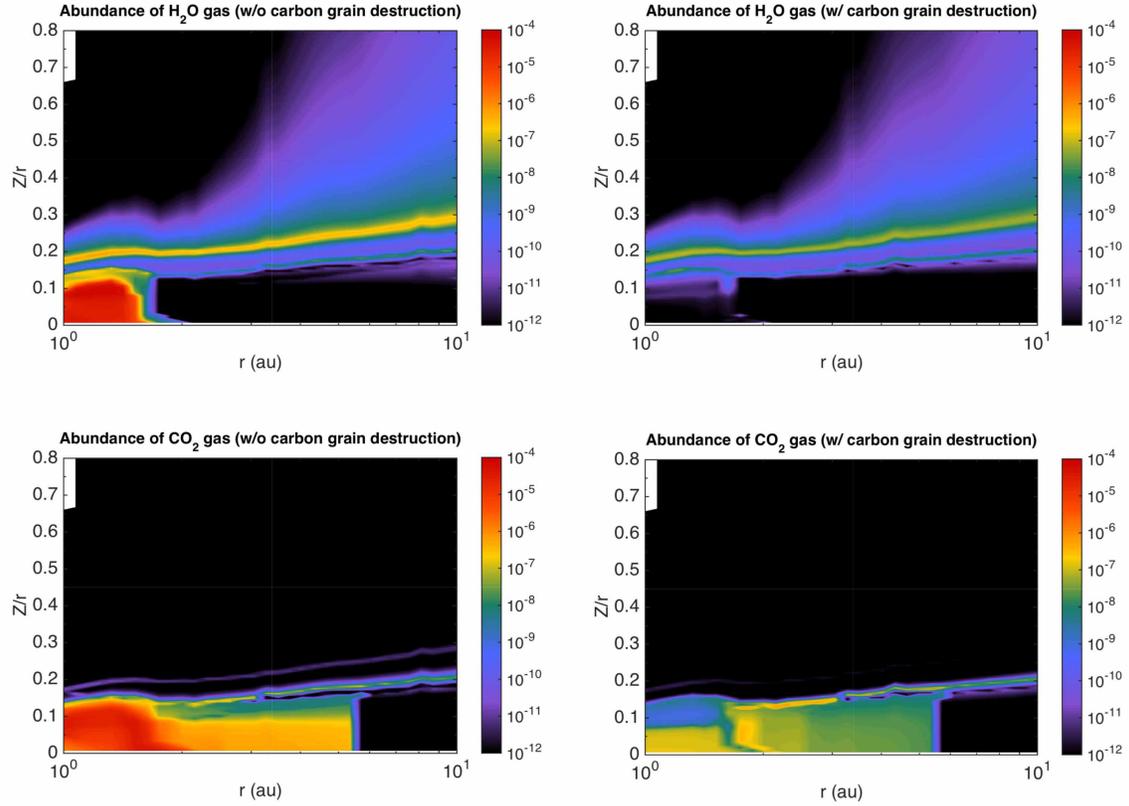}}
\caption{The 2-dimensional distribution of fractional abundances of oxygen-bearing molecules in the gas-phase for the models without (left) and with (right) carbon grain destruction as a function of radius and height of the disk for a T Tauri disk.}
\label{fig:11}
\end{figure}

\begin{figure} [h]
\centerline{\includegraphics[width=0.7\textwidth]{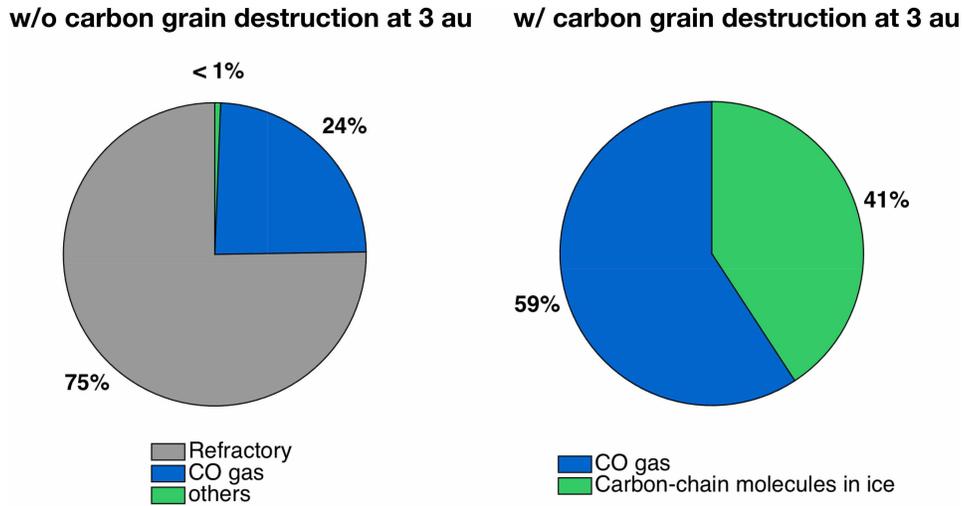}}
\caption{Pie chart showing the percentage of the form of carbon (gas and solid) for the models without (left) and with (right) carbon grain destruction at a disk radius of 3 au for a T Tauri disk.}
\label{fig:12}
\end{figure}

\begin{figure} [h]
\centerline{\includegraphics[width=0.8\textwidth]{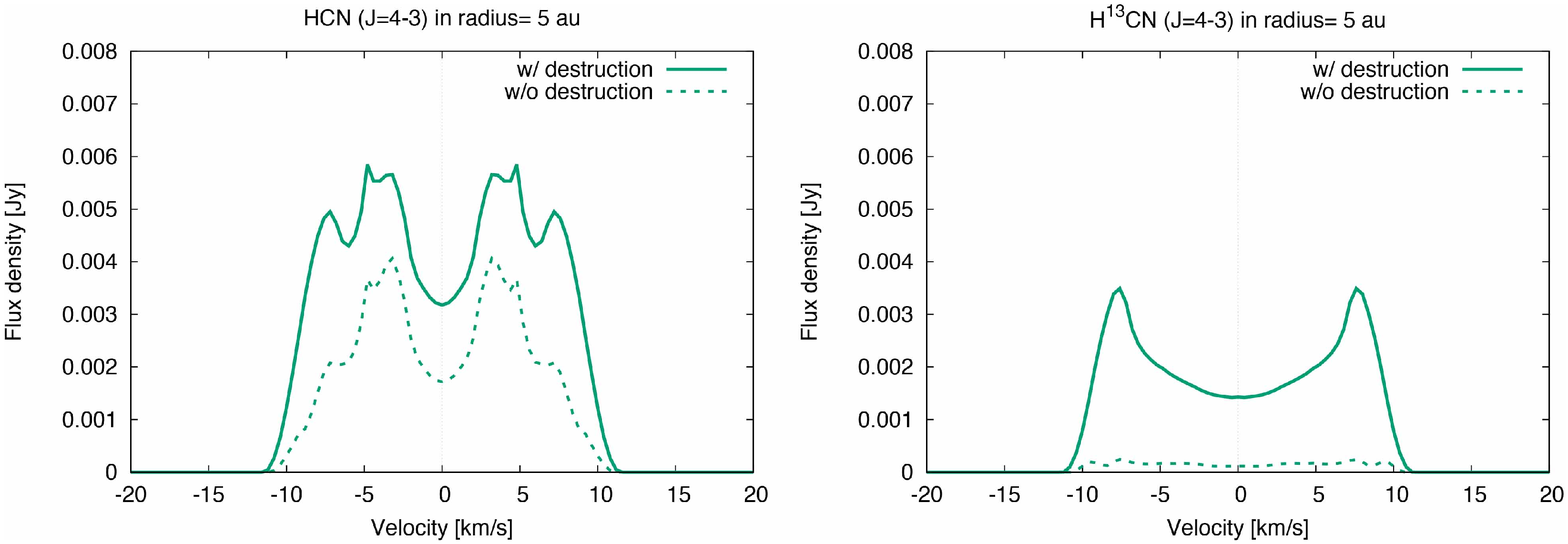}}
\caption{ The line profiles of the HCN (left) and \ce{H^{13}CN} (right) J = 4 - 3 lines within r $<$ 5 au of T-Tauri disk. Solid and dashed lines represent the models with and without carbon grain destruction, respectively. }
\label{fig:13}
\end{figure}

\begin{figure} [h]
\centerline{\includegraphics[width=1\textwidth]{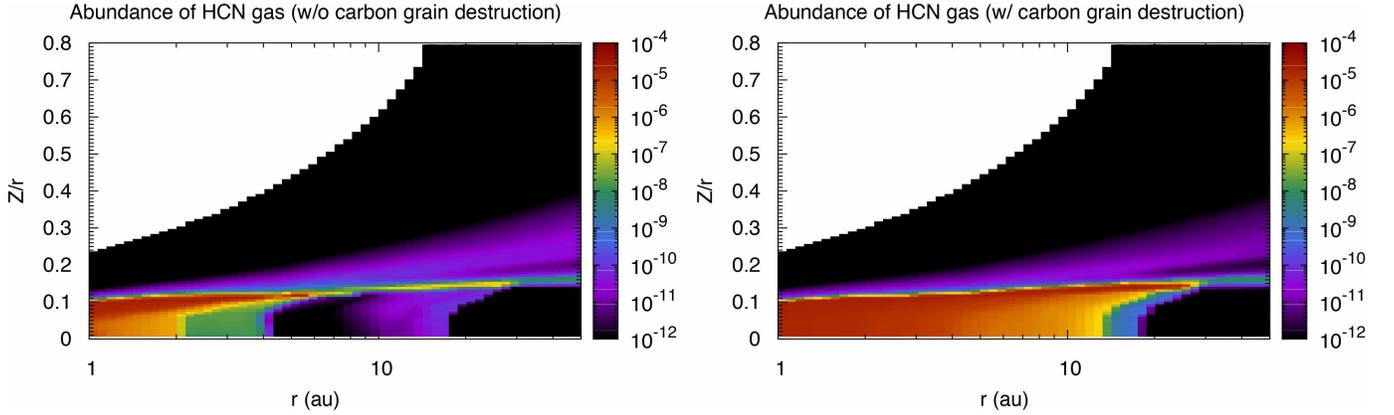}}
\caption{ The abundance distributions of HCN in a protoplanetary disk  around a Herbig Ae star for models without (left-hand-side) and with (right-hand-side) carbon grain destruction. A dramatic change appears around r $\sim$ 2-30 au between the models.}
\label{fig:14}
\end{figure}

\begin{figure} [h]
\centerline{\includegraphics[width=0.7\textwidth]{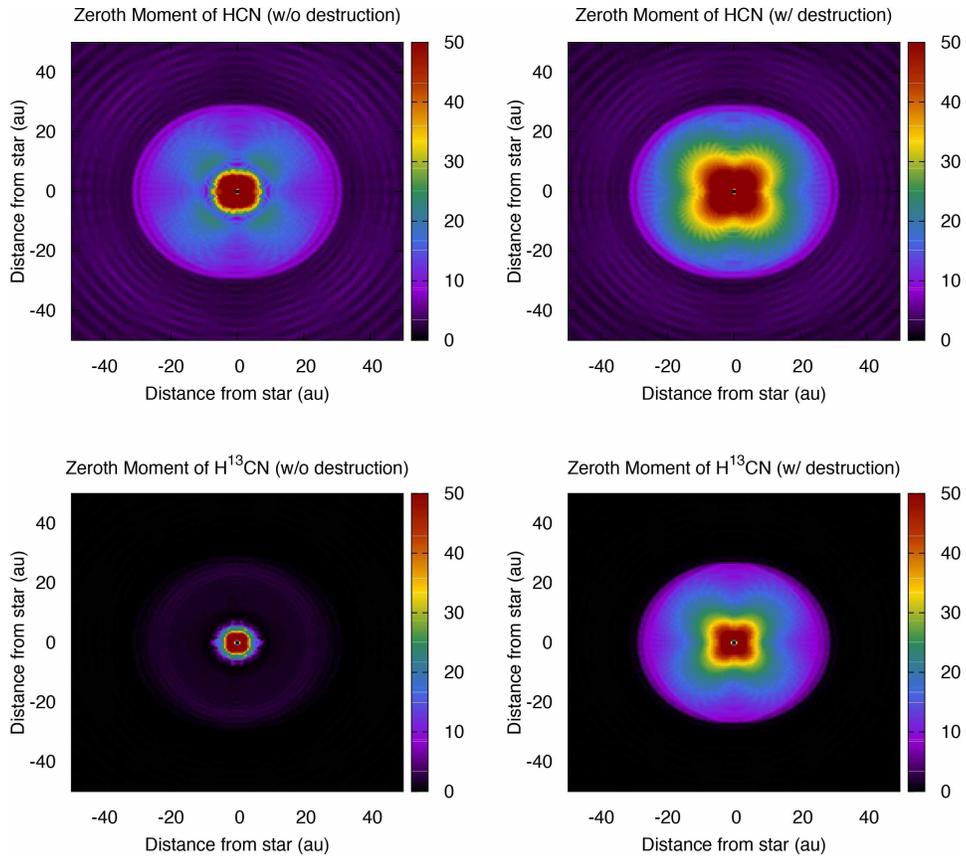}}
\caption{ The zeroth moment maps of the HCN (top) and \ce{H^{13}CN} (bottom) J= 4 -3 lines for the models without (left-hand-side) and with (right-hand-side) the carbon grain destruction for the Herbig Ae disk. The unit of the color map is Jy arcsec$^{-2}$km s$^{-1}$.}
\label{fig:15}
\end{figure}

\begin{figure} [h]
\centerline{\includegraphics[width=0.7\textwidth]{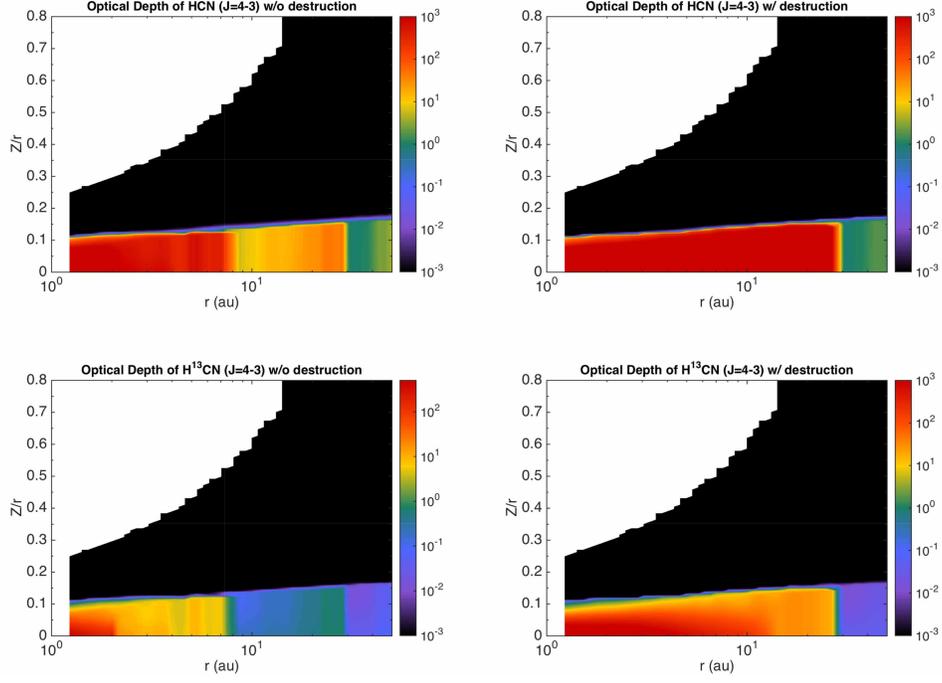}}
\caption{The optical depth of the HCN (top) and \ce{H^{13}CN} (bottom) lines for the models without (left-hand-side) and with (right-hand-side) the carbon grain destruction.}
\label{fig:16}
\end{figure}

\begin{figure}
\centerline{\includegraphics[width=0.8\textwidth]{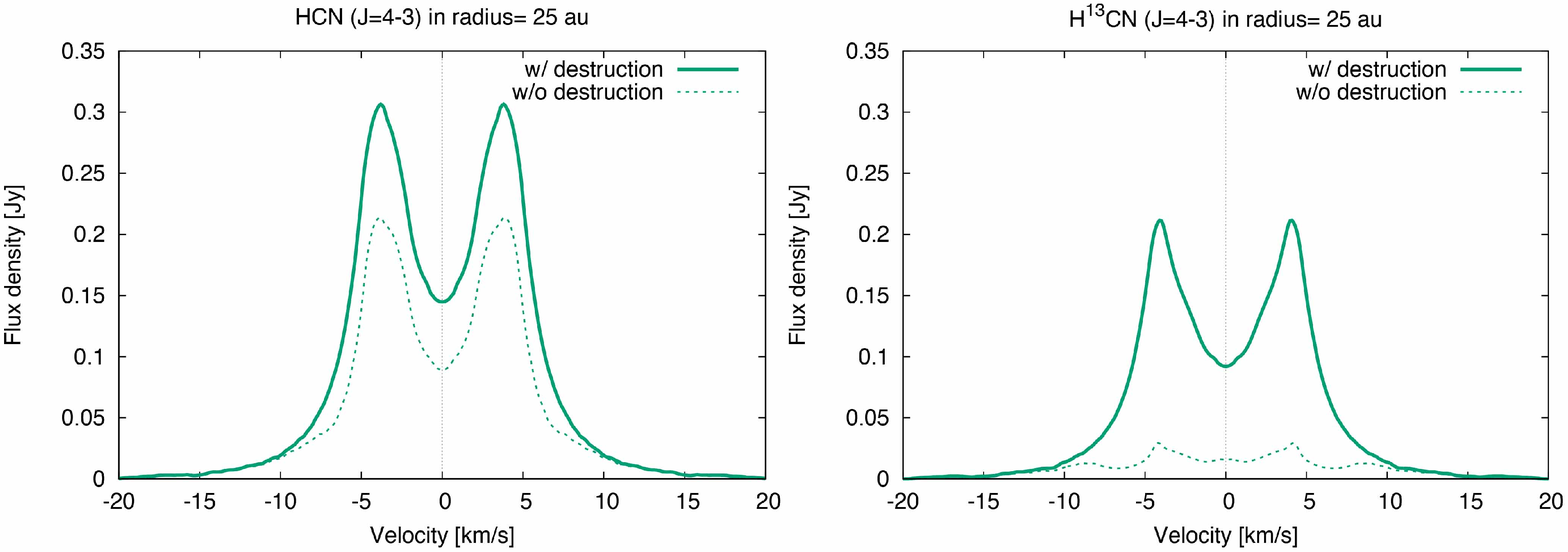}}
\caption{ The line profiles of the HCN (left) and \ce{H^{13}CN} (right) J = 4 - 3 lines within r $<$ 25 au of Herbig Ae disk. Solid and dashed lines represent the models with and without carbon grain destruction, respectively. }
\label{fig:17}
\end{figure}

\begin{figure} 
\centerline{\includegraphics[width=0.7\textwidth]{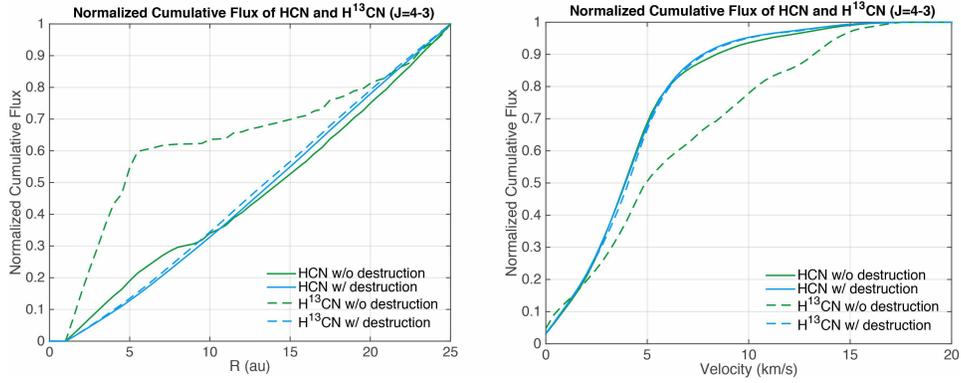}}
\caption{ The normalized cumulative line flux of the HCN (solid lines) and \ce{H^{13}CN} (dashed lines) J = 4 - 3 lines as a function of the disk radius (left-hand-side) and the velocity (right-hand-side) for the models without (green lines) and with (blue lines) carbon grain destruction. Under the assumption of a disk in Keplerian rotation, the velocity profiles reflect the line emitting regions.}
\label{fig:18}
\end{figure}

\begin{figure} 
\centerline{\includegraphics[width=1\textwidth]{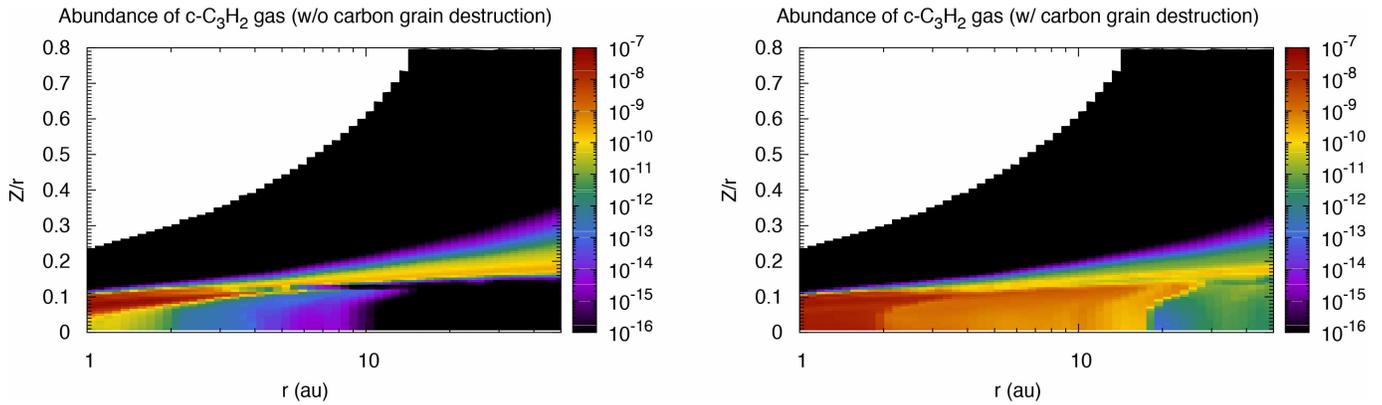}}
\caption{The abundance distributions of \ce{c-C_3H_2} in a protoplanetary disk  around a Herbig Ae star for models without (left-hand-side) and with (right-hand-side) carbon grain destruction.}
\label{fig:19}
\end{figure}

\begin{figure} 
\centerline{\includegraphics[width=0.8\textwidth]{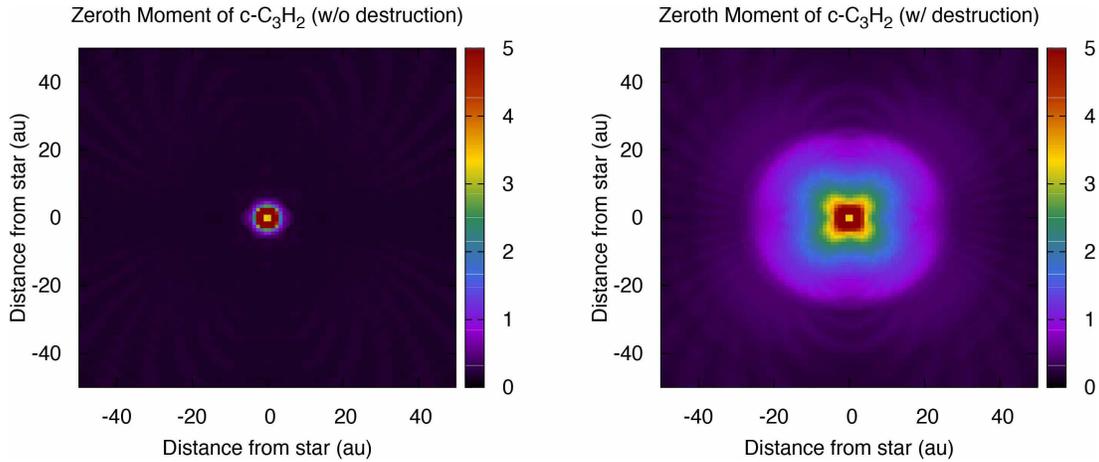}}
\caption{The same as Figure 15 but for the c-\ce{C3H2} $6_{1,6}$-$5_{0,5}$ line for the Herbig Ae disk.}
\label{fig:20}
\end{figure}

\begin{figure} 
\centerline{\includegraphics[width=0.4\textwidth]{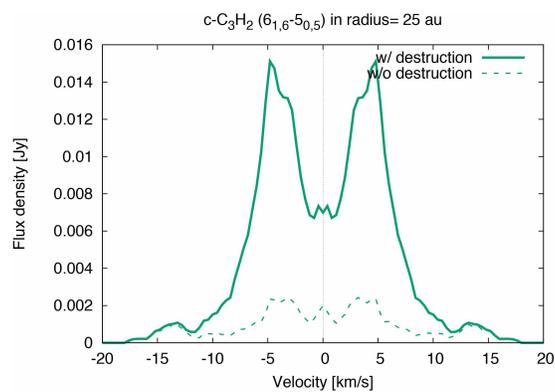}}
\caption{ The same as Figure 17 but for the c-\ce{C3H2} $6_{1,6}$-$5_{0,5}$ line within r $<$ 25 au.}
\label{fig:21}
\end{figure}



\end{document}